\RequirePackage{fix-cm}
\documentclass[10pt]{article}         
\usepackage{amsmath}\usepackage{amsmath,amsfonts,amssymb,amsthm,url,array}
\usepackage{float}
\usepackage{mathbbol}
\usepackage{abstract}
\usepackage{hyperref}
\usepackage[utf8]{inputenc}
\usepackage[T1]{fontenc}

\usepackage[sort]{natbib}

\usepackage{setspace}
\usepackage{mathptmx}
\usepackage{titlesec}
\titleformat{\paragraph}
{\normalfont\normalsize\bfseries}{\theparagraph}{1em}{}
\titlespacing*{\paragraph}
{0pt}{3.25ex plus 1ex minus .2ex}{1.5ex plus .2ex}
\usepackage{optidef}
\usepackage{mathrsfs} 
\usepackage{graphicx}
\allowdisplaybreaks
\newtheorem{theorem}{Theorem}

\newtheorem{prop}{Proposition}
\newtheorem{Results}{Result}

\usepackage{optidef}
\usepackage{graphicx}
\usepackage{bbm}
\usepackage{mathtools}
\usepackage{graphicx}
\usepackage{caption}
\usepackage{subcaption}
\allowdisplaybreaks
\begin{document}

\title{Multi-asset Generalised Variance Swaps \\ in Barndorff-Nielsen and Shephard model}

\author{
    Subhojit Biswas\\
    \textit{Indian Statistical Institute, Kolkata, India}\\
    \textit{subhojit1016kgp@gmail.com}\\
    \\
    Diganta Mukherjee\\
    \textit{Sampling and Official Statistics Unit}\\
    \textit{Indian Statistical Institute, Kolkata, India}\\
    \textit{digantam@hotmail.com}\\
    \\
    Indranil SenGupta\\
    \textit{Department of Mathematics}\\
    \textit{North Dakota State University, Fargo, North Dakota, USA}\\
    \textit{indranil.sengupta@ndsu.edu}
}

\maketitle

\begin{abstract}
This paper proposes swaps on two important new measures of generalized variance,  namely the maximum eigenvalue and trace of the covariance matrix of the assets involved. We price these generalized variance swaps for Barndorff-Nielsen and Shephard model used in financial markets. We consider multiple assets in the portfolio for  theoretical purpose and demonstrate our approach with numerical examples taking three stocks in the portfolio. The results obtained in this paper have important implications for the commodity sector where such swaps would be useful for hedging risk. 
\end{abstract} 



\thispagestyle{empty}
\newpage

\setcounter{page}{1}

\section{Introduction}

Covariance and correlation swaps are among recent financial products which are useful for volatility hedging and speculation using two different financial underlying assets. For example, option dependent on exchange rate movements, such as those paying in a currency different from the underlying currency, have an exposure to movements of the correlation between the asset and the exchange rate, this risk may be eliminated by using a covariance swap.

The literature devoted to the volatility derivatives is growing. The Non-Gaussian Ornstein-Uhlenbeck stochastic volatility model was used by \citeauthor{benth} (\citeyear{benth}) to study volatility and variance swaps. \citeauthor{broadie1} (\citeyear{broadie1}) evaluated price and hedging strategy for volatility
derivatives in the Heston square root stochastic volatility model and in \citeauthor{broadie2} (\citeyear{broadie2}) they compare result from various model in order to investigate the effect of jumps and discrete sampling on variance and volatility swaps. Pure jump process with independent increments return models were used by \citeauthor{carr1}(\citeyear{carr1}) to price derivatives written on realized variance, and
subsequent development by \citeauthor{carr1}(\citeyear{carr1}). This paper also provides a good survey on volatility derivatives. \citeauthor{fonseca}(\citeyear{fonseca}) analyzed the influence of variance and covariance swap in a market by solving a portfolio optimization problem in a market with risky assets and volatility derivatives. Correlation swap price has been investigated by \citeauthor{bossu1} (\citeyear{bossu1}) and \citeauthor{bossu2} (\citeyear{bossu2}) for component of an equity index using statistical method. \citeauthor{sengupta} (\citeyear{sengupta}) developed the covariance swap price for a pair of assets using the  well known Barndorff-Nielsen and Shephard (BNS) model.

By definition, all the covariance based methods discussed above can only consider a combination of two assets at a time. But in today's complex financial transactions, there is no reason why volatility of three or more assets will not be considered for contracting together. To wit, over the last few decades, financial industry has introduced various products that are directly linked to volatility for the purposes of indexing and hedging volatility risk. Such financial products include variance swap, volatility swap, option on realized variance, etc. CBOE introduced the Volatility Index (symbol VIX), as a reference for the 30-day volatility of the S\&P500. It launched the trading of futures on VIX in 2004 and, later in 2006, the trading of options on VIX. Subsequently CBOE also published “VIX” on other indices such as VXN (volatility index of NASDAQ), VXD (volatility of DJIA), and RVX (volatility of Russell 2000), etc. (See \citeauthor{survey} (\citeyear{survey}) for a detailed discussion.)
In the presence of such multiple volatility linked indices and contracts, where the underlying assets are closely interlinked, it very much makes sense to consider a measure of (multivariate) volatility of these underliers together as a vector.

Thus, in this paper, we extend these methods to a situation where a suitably defined generalized variance of a portfolio of assets can be contracted on. Taking cue from multivariate analysis, we look at two important measures of generalized variance, namely the {\it maximum eigenvalue} and {\it trace} of the covariance matrix of the assets involved. \citeauthor{Ref3}(\citeyear{Ref3}) develops the theory for pricing such assets for a Markov-modulated volatility situation. 
The objective of the present paper is to extend this analysis to price generalized variance swaps for financial markets with jumps in the return process. Our candidate model for the returns is the well known BNS model.

We outline the problem and the theoretical results in section 2. The application to the pricing of Swaps is done in Section 3. First we work out the price of the trace swap and in a subsequent subsection we discuss pricing of the eigenvalue swap with a target return constraint. The numerical examples are presented with real data in section 4. Finally section 5 concludes.

\section{Problem formulation}

Let us consider a financial market with two types of securities, the risk free bond and the stock. Barndorff-Nielsen and Shephard assumed that the price process of the stock $ S = (S_t)_{t\geq 0}$ is defined on some filtered probability space $(\Omega, \mathscr{F}, (\mathscr{F}_t)_{0 \leq t \leq T}, P)$ and is given by 
\begin{equation}
S_t = S_0 e^{X_t}. \nonumber
\end{equation}
\begin{equation}
d X_t = (\mu + \beta \sigma^2_t) dt + \sigma_t dW_t + \rho dZ_{\lambda t}. \nonumber
\end{equation}
\begin{equation}
d \sigma^2_t = -\lambda \sigma^2_t dt + dZ_{\lambda t} , \sigma^2_{0}\geq 0 .\nonumber
\end{equation}

where the parameters $\mu, \beta, \rho, \lambda \in R $ with $\lambda > 0$ and $\rho \leq 0$. $W = (W_t)$ is a Brownian motion and the process $Z = (Z_{\lambda t})$ is a subordinator. Barndorff-Nielsen and Shephard refer to Z as the background driving L\'evy process (BDLP). Z satisfies the Assumptions [1 - 3] in the paper \citeauthor{sengupta} (\citeyear{sengupta}) that there exists an equivalent martingale
measure under which the equations can be written as:

\begin{equation}
d X_t = b_t dt + \sigma_t dW_t + \rho dZ_{\lambda t}. \nonumber
\end{equation}
\begin{equation}\label{eq1}
d \sigma^2_t = -\lambda \sigma^2_t dt + dZ_{\lambda t} , \sigma^2_{0}\geq 0.
\end{equation}
\begin{equation}
    b_t = (r - \lambda \kappa (\rho) - \frac{1}{2} \sigma^2_t). \nonumber
\end{equation}
and $W_t$ and $Z_{\lambda t}$ are Brownian motion and L\'evy process respectively with respect to
the equivalent martingale measure. In the expression for $b_t$, the cumulant transform
for $Z_1$ under the new measure is denoted as $\kappa (\theta)$. The solution of the \eqref{eq1} is given by
\begin{equation}
    \sigma^2_t = e^{-\lambda t}\sigma^2_{0} + \int_{0}^{t}e^{-\lambda (t -s)}d Z_{\lambda s}. \nonumber
\end{equation}
Here Z is an increasing function and $\sigma^2_{0} > 0$, the process $\sigma^2 = (\sigma^2_t)$ is strictly positive and is bounded by $\sigma^2_{0}e^{-\lambda t}$. The instantaneous variance of log returns is given by $(\sigma^2_t + \rho^2 \lambda Var[Z_1]dt)$ and therefore simple calculation shows the continuous realized variance in the interval [0, T] is
\begin{equation}
    \sigma^2_R = \frac{1}{T} \int_{0}^{T} \sigma^2_t dt+ \rho^2 \lambda Var[Z_1]. \nonumber
\end{equation}
Substituting $\sigma^2_t$ to the realized variance $\sigma^2_R$ and using integration by parts
of Ito calculus yield
\begin{equation}
    \sigma^2_R = \frac{1}{T}\bigg(\lambda^{-1}(1-e^{-\lambda T})\sigma^2_{0} + \lambda^{-1}\int_{0}^{T}(1 - e^{-\lambda (T - s)})d Z_{\lambda s}\bigg) + \rho^2 \lambda Var[Z_1]. \nonumber
\end{equation}

A portfolio consists of 3 stocks with the corresponding returns given by $dX^1$, $dX^2$ and $dX^3$. The risk-neutral dynamics of the assets are given as:
\begin{equation}\label{eq2}
d X_t^{i} = b_t^{i} dt + (\sigma_{i})_t dW^{i}_t + \rho_{i} dZ^{1}_{\lambda t}. 
\end{equation}
where $i = \{1,2,3\}$
\begin{equation}
d (\sigma_{1})^2_t = -\lambda (\sigma_{1})^2_t dt + dZ^{1}_{\lambda t}, \quad (\sigma_{1})^2_{0}\geq 0 \nonumber
\end{equation}
\begin{equation}
d (\sigma_{2})^2_t = -\lambda (\sigma_{2})^2_t dt + dZ^{2}_{\lambda t}, \quad (\sigma_{2})^2_{0}\geq 0 \nonumber
\end{equation}
\begin{equation}\label{eq3}
d (\sigma_{3})^2_t = -\lambda (\sigma_{3})^2_t dt + dZ^{3}_{\lambda t}, \quad  (\sigma_{3})^2_{0}\geq 0 
\end{equation}
where $W^{i}_t$'s are the Wiener processes and correlated with other Wiener process as Cov$(W^{i}_t, W^{j}_t) = \gamma_{ij}t$ and $b^{i}_t$'s are deterministic functions of $(\sigma_{i})^2_t$. Here $\rho_{i}$'s are the leverage parameter corresponding to $S^{i}$'s respectively.\newline
\begin{prop}
(\citeauthor{sengupta}(\citeyear{sengupta})): Let $(Z^{1}_{t}, Z^{*}_{t})$, $(Z^{*}_{t}, Z^{**}_{t})$ and $(Z^{1}_{t}, Z^{**}_{t})$  be three pairs of independent L\'evy subordinators. Here the independence of the L\'evy processes is understood in the sense of \citeauthor{Cont} (\citeyear{Cont}) (Proposition 5.3). If $(X_t, Y_t)$ is a L\'evy process with L\'evy measure $\nu_{(X, Y)}$ and without Gaussian part then its components are independent if and only if the support of $\nu_{(X, Y)}$ is contained in the set $\{(x, y) : xy = 0\}$, that is, if and only if they never jump together with probability one.
In this case $\nu_{(X, Y)}(A) = \nu_X(A_X) + \nu_Y(A_Y)$, where $A_X = \{x : (x, 0) \in A\}$ and
$A_Y = \{y : (y, 0) \in A\}$, and $\nu_X$ and $\nu_Y$ are L\'evy measures of $X_t$ and $Y_t$. So we define,
\begin{equation}\label{eq4}
    d Z^{2}_t = r_{2}d Z^{1}_t + \sqrt{1 - (r_{2})^2} d Z^{*}_{t}.
\end{equation}
\begin{equation}\label{eq5}
    d Z^{3}_t = r_{3}d Z^{1}_t + \sqrt{1 - (r_{3})^2} d Z^{**}_{t}.
\end{equation}
From the above two equations we can write,
\begin{equation}
r_{3} d Z^{2}_t - r_{2}d Z^{3}_t = r_{3}\sqrt{1 - (r_{2})^2} d Z^{*}_{t} - r_{2}\sqrt{1 - (r_{3})^2} d Z^{**}_{t}, \nonumber
\end{equation}
\begin{equation} \label{eq6}
 d Z^{3}_t = \frac{r_{3}}{r_{2}} d Z^{2}_t  - \frac{r_{3}}{r_{2}} \sqrt{1 - (r_{2})^2} d Z^{*}_{t} + \sqrt{1 - (r_{3})^2} d Z^{**}_{t}. 
\end{equation}

which is also a L\'evy subordinator provided $0 \leq r_{2}, r_{3} \leq 1$. 
\end{prop}
Thus, for $0 \leq r_{2}, r_{3} \leq 1$, $(Z^{1}_{t}, Z^{2}_{t})$, $(Z^{1}_{t}, Z^{3}_{t})$ and $(Z^{2}_{t}, Z^{3}_{t})$ are positively correlated L\'evy subordinators \citeauthor{Ref1}(\citeyear{Ref1}). It is clear that $Var(Z^{1}_t) = t\kappa^{1}_2$, $Var(Z^{2}_t) = t\kappa^{2}_2$, $Cov(Z^{1}_t, Z^{2}_t) = r_{2} Var(Z^{1}_t) = r_{2}t\kappa^{1}_2$, where $\kappa^{1}_2$ and $\kappa^{2}_2$ are the variances (second cumulant) of $Z^{1}_t$ and $Z^{2}_t$ respectively. Therefore the correlation coefficient between $Z^{1}_t$ and $Z^{2}_t$ at any t is given by $r_{2}\sqrt{\frac{\kappa^{1}_2}{\kappa^{2}_2}}$. Similarly, the correlation coefficient between $Z^{3}_t$ and $Z^{1}_t$ at any t is given by $r_{3}\sqrt{\frac{\kappa^{1}_2}{\kappa^{3}_2}}$ and the correlation coefficient between $Z^{2}_t$ and $Z^{3}_t$ at any t is given by $\frac{r_{3}r_{2}\kappa^{1}_2}{\sqrt{\kappa^{2}_2\kappa^{3}_2}}$. \newline

Notation: $[\cdot , \cdot]$ represents the quadratic covariation. \newline

Let the portfolio return covariance matrix be given by
\[ \Omega = 
\begin{bmatrix}
    Cov_R(S^1, S^1) & Cov_R(S^1, S^2) & Cov_R(S^1, S^3)  \\
    Cov_R(S^2, S^1) & Cov_R(S^2, S^2) & Cov_R(S^2, S^3) \\ Cov_R(S^3, S^1) & Cov_R(S^3, S^2) & Cov_R(S^3, S^3)  
\end{bmatrix}
\]
From the definition of $Cov_R(S^{1}, S^{2})$ we have
\begin{equation}
    Cov (S^{1}, S^{2}) = \frac{1}{T} [\ln S^{1}_T, \ln S^{2}_T] = \frac{1}{T}[X^{1}_T, X^{2}_T] = Cov (S^{2}, S^{1}).  \nonumber
\end{equation}
Similarly we can write, 
\begin{equation}
    Cov (S^{2}, S^{3}) = \frac{1}{T} [\ln S^{2}_T, \ln S^{3}_T] = \frac{1}{T}[X^{2}_T, X^{3}_T] = Cov (S^{3}, S^{2}).  \nonumber
\end{equation}
\begin{equation}
    Cov (S^{1}, S^{3}) = \frac{1}{T} [\ln S^{1}_T, \ln S^{3}_T] = \frac{1}{T}[X^{1}_T, X^{3}_T] = Cov (S^{3}, S^{1}).  \nonumber
\end{equation}
Considering the Assumption 4 and Remark (3.2) from \citeauthor{sengupta} (\citeyear{sengupta}) we can write the solution of Equation \eqref{eq2} as
\begin{equation}
X_t^{i} = \int_{0}^{T}b_t^{i} dt +\int_{0}^{T} (\sigma_{i})_t dW^{i}_t + \int_{0}^{\lambda T} \int_{0}^{\infty}\rho_{i} y J_{Z^1}(ds, dy).  \nonumber
\end{equation}
Therefore $[X^{1}_T, X^{2}_T]$ is given by \citeauthor{Cont} (\citeyear{Cont}) (Section 8.2.2)
\begin{equation}
    [X^{1}_T, X^{2}_T] = \int_{0}^{T} \gamma_{12} (\sigma_{1})_t (\sigma_{2})_t dt +  \rho_{1}\rho_{2}\int_{0}^{\lambda T} \int_{0}^{\infty} y^2 J_{Z^1}(ds, dy). \nonumber
\end{equation}
Similarly,
\begin{equation}
    [X^{2}_T, X^{3}_T] = \int_{0}^{T} \gamma_{23} (\sigma_{2})_t (\sigma_{3})_t dt +  \rho_{2}\rho_{3}\int_{0}^{\lambda T} \int_{0}^{\infty} y^2 J_{Z^1}(ds, dy). \nonumber
\end{equation}
\begin{equation}\label{eq7}
    [X^{1}_T, X^{3}_T] = \int_{0}^{T} \gamma_{13} (\sigma_{1})_t (\sigma_{3})_t dt +  \rho_{1}\rho_{3}\int_{0}^{\lambda T} \int_{0}^{\infty} y^2 J_{Z^1}(ds, dy).
\end{equation}

We will now try to find the value for 
\begin{equation}
 E [Cov(S^{1},S^{2})] = \frac{1}{T}E[X^{1}_T, X^{2}_T].  \nonumber 
\end{equation}
\begin{equation}
 E [Cov(S^{2},S^{3})] = \frac{1}{T}E[X^{2}_T, X^{3}_T].  \nonumber 
\end{equation}
\begin{equation}
 E [Cov(S^{3},S^{1})] = \frac{1}{T}E[X^{3}_T, X^{1}_T].  \nonumber 
\end{equation}

Stating Lemma 3.3 from \citeauthor{sengupta}(\citeyear{sengupta}) as a theorem
\begin{theorem}
If $B_t$ has been defined as
\begin{equation}\label{eq15}
    B_t = \alpha_1 + \int_{0}^{\lambda T} e^{s}d V_s.
\end{equation}
where $\lambda> 0$, $\alpha_1, \lambda \in R$ are constants and $0 \leq t \leq T$ and V is a L\'evy subordinator with no deterministic drift. Then,
\begin{equation}
    \phi_{B_{t}}(\theta) = exp \bigg(i\theta \alpha_1 + \int_{0}^{\lambda t} \kappa_{V_{1}}\Big(i\theta e^{s}\Big)ds\bigg). \nonumber
\end{equation}
where $\kappa_{V_{1}}(\cdot)$ is the cumulant generating function for $V_1$. The moment of $B_t$ are given by
\begin{equation}
    E(B^{k}_t) = (-i)^{k}\tilde{g_k} (0), \quad k = 1,2,\dots \nonumber
\end{equation}
where
\begin{equation}
    \tilde{g_1} (\theta) = i \bigg(\alpha_1 + \int_{0}^{\lambda t}e^{s} \kappa'_{V_{1}}\Big(i\theta e^{s}\Big)ds\bigg), \nonumber
\end{equation}
and
\begin{equation}
    \tilde{g_{k+1}} (\theta) = \tilde{g_1} (\theta)\tilde{g_k} (\theta) + \tilde{g_k}' (\theta), \quad k = 1,2,\dots \nonumber
\end{equation}
In the above formulas prime represents the derivative with respect to the parameter
in parenthesis.
\end{theorem}

\begin{theorem}
The correlation coefficient between $(\sigma_{2})^2_t$ \& $(\sigma_{3})^2_t$, $(\sigma_{2})^2_t$ \& $(\sigma_{3})^2_t$ and $(\sigma_{2})^2_t$ \& $(\sigma_{3})^2_t$ can be calculated as;

\begin{enumerate}
\item Suppose $Z^{2}$, $Z^{3}$, $Z^{*}$  and $Z^{**}$ are related by \eqref{eq6} then the correlation coefficient between $(\sigma_{2})^2_t$ and $(\sigma_{3})^2_t$ is independent of time and is given by $\frac{r_{3}r_{2}\kappa^{1}_2}{\sqrt{\kappa^{2}_2\kappa^{3}_2}}$ where $\kappa^{1}_2$, $\kappa^{2}_2$ and $\kappa^{3}_2$ are the variances (second cumulant) of $Z^{1}_t$, $Z^{2}_t$ and $Z^{3}_t$ respectively.

\item Suppose $Z^{2}$, $Z^{1}$ and $Z^{1*}$ are related by \eqref{eq4} then the correlation coefficient between $(\sigma_{1})^2_t$ and $(\sigma_{2})^2_t$ is independent of time and is given by $r_{2}\sqrt{\frac{\kappa^{1}_2}{\kappa^{2}_2}}$ where $\kappa^{1}_2$ and $\kappa^{2}_2$ are the variances (second cumulant) of $Z^{1}_t$ and $Z^{2}_t$ respectively.\newline

\item Suppose $Z^{3}$, $Z^{1}$ and $Z^{**}$ are related by \eqref{eq5} then the correlation coefficient between $(\sigma_{1})^2_t$ and $(\sigma_{3})^2_t$ is independent of time and is given by $r_{3}\sqrt{\frac{\kappa^{1}_2}{\kappa^{3}_2}}$ where $\kappa^{1}_2$ and $\kappa^{3}_2$ are the variances (second cumulant) of $Z^{1}_t$ and $Z^{3}_t$ respectively.

\end{enumerate}
\end{theorem}

\textit{Proof:} It is clear from Equation \eqref{eq3} that,

\begin{equation}
    (\sigma_2)^{2}_t = e^{-\lambda t}(\sigma_2)^{2}_0 + e^{-\lambda t} \int_{0}^{ t} e^{\lambda s} d Z^{2}_{\lambda s}. \nonumber
\end{equation}
\begin{equation}
    (\sigma_3)^{2}_t = e^{-\lambda t}(\sigma_3)^{2}_0 + e^{-\lambda t} \int_{0}^{ t} e^{\lambda s} d Z^{3}_{\lambda s}. \nonumber
\end{equation}
Using Theorem 1 we get,
\begin{equation}
    Var((\sigma_2)^{2}_t) = e^{-2\lambda t}Var\bigg(\int_{0}^{ t} e^{\lambda s} d Z^{2}_{\lambda s}\bigg) = e^{-2\lambda t}\frac{k^2_2}{2}(e^{2\lambda t} - 1).\nonumber
\end{equation}
Similarly,

\begin{equation}
    Var((\sigma_3)^{2}_t) = e^{-2\lambda t}Var\bigg(\int_{0}^{ t} e^{\lambda s} d Z^{3}_{\lambda s}\bigg) = e^{-2\lambda t}\frac{k^3_2}{2}(e^{2\lambda t} - 1).\nonumber
\end{equation}
\begin{equation}
    (\sigma_2)^{2}_t = e^{-\lambda t}(\sigma_2)^{2}_0 + e^{-\lambda t} \int_{0}^{ t} e^{\lambda s} d Z^{2}_{\lambda s} = e^{-\lambda t}(\sigma_2)^{2}_0 + e^{-\lambda t}\int_{0}^{ t} e^{\lambda s}(r_2 d Z^{1}_{\lambda s} + \sqrt{1 - r^2_2} d Z^{*}_{\lambda s}),\nonumber
\end{equation}
\begin{equation}
    (\sigma_2)^{2}_t = r_2(\sigma_{1})^2_t + e^{-\lambda t}((\sigma_2)^{2}_0 - r_2(\sigma_{1})^2_0) +   \sqrt{1 - r^2_2} \int_{0}^{ t}  d Z^{*}_{\lambda s}.\nonumber
\end{equation}
\begin{equation}
    (\sigma_3)^{2}_t = r_3(\sigma_{1})^2_t + e^{-\lambda t}((\sigma_3)^{2}_0 - r_3(\sigma_{1})^2_0) +   \sqrt{1 - r^2_3} \int_{0}^{ t}  d Z^{**}_{\lambda s}.\nonumber
\end{equation}
Now,
\begin{equation}
    Cov((\sigma_2)^{2}_t, (\sigma_3)^{2}_t) = E\Big((\sigma_2)^{2}_t (\sigma_3)^{2}_t\Big) - E\Big((\sigma_2)^{2}_t\Big)E\Big((\sigma_3)^{2}_t\Big),\nonumber
\end{equation}
\begin{eqnarray}
    &=& E\bigg(\Big(r_2(\sigma_{1})^2_t + e^{-\lambda t}((\sigma_2)^{2}_0 - r_2(\sigma_{1})^2_0) +   \sqrt{1 - r^2_2} \int_{0}^{ t}  d Z^{*}_{\lambda s}\Big)\Big(r_3(\sigma_{1})^2_t + e^{-\lambda t}((\sigma_3)^{2}_0 - r_3(\sigma_{1})^2_0) +   \sqrt{1 - r^2_3} \int_{0}^{ t}  d Z^{**}_{\lambda s}\Big)\bigg) - \nonumber \\&& E\Big(r_2(\sigma_{1})^2_t + e^{-\lambda t}((\sigma_2)^{2}_0 - r_2(\sigma_{1})^2_0) +   \sqrt{1 - r^2_2} \int_{0}^{ t}  d Z^{*}_{\lambda s}\Big)E\Big(r_3(\sigma_{1})^2_t + e^{-\lambda t}((\sigma_3)^{2}_0 - r_3(\sigma_{1})^2_0) +   \sqrt{1 - r^2_3} \int_{0}^{ t}  d Z^{**}_{\lambda s}\Big),\nonumber
\end{eqnarray}
Simplifying it,
\begin{equation}
    Cov((\sigma_2)^{2}_t, (\sigma_3)^{2}_t) = r_2r_3E\Big((\sigma_1)^{4}_t) - r_2r_3 E\Big((\sigma_1)^{2}_t\Big)E\Big((\sigma_1)^{2}_t\Big) = r_2r_3\bigg(E\Big((\sigma_1)^{4}_t)\Big) - E\Big((\sigma_1)^{2}_t\Big)E\Big((\sigma_1)^{2}_t\Big)\bigg) = r_2r_3Var\Big((\sigma_1)^{2}_t\Big), \nonumber
\end{equation}
\begin{equation}
    Cov((\sigma_2)^{2}_t, (\sigma_3)^{2}_t) = r_2r_3 e^{-2\lambda t}\frac{k^1_2}{2}(e^{2\lambda t} - 1).\nonumber
\end{equation}
\begin{equation}
    Correlation = \frac{Cov((\sigma_2)^{2}_t, (\sigma_3)^{2}_t)}{\sqrt{Var((\sigma_2)^{2}_t)Var((\sigma_3)^{2}_t)}} = \frac{r_{3}r_{2}\kappa^{1}_2}{\sqrt{\kappa^{2}_2\kappa^{3}_2}}. \nonumber
\end{equation}

We can similarly prove the second and third part of the Theorem.
Hence Theorem 2 is proved. $\qed$ \newline

\begin{theorem}
(Lemma 3.5 from  \citeauthor{sengupta}(\citeyear{sengupta}))\begin{enumerate}\item Suppose that $(\sigma_{1})^2_t$ and $(\sigma_{2})^2_t$ 
defined respectively by \eqref{eq3} and $Z^{1}$, $Z^{2}$ and $Z^{*}$ are related by \eqref{eq4}. Let there exists $\beta_{12} >0$ such that for $0 \leq t \leq T$, $(\sigma_{1})^2_t, (\sigma_{2})^2_t \leq \beta^2_{12}$. Then,
\begin{eqnarray}\label{eq8}
    E[(\sigma_{1})_t(\sigma_{2})_t] = e^{-\lambda t}\sum_{k=0}^{\infty}\sum_{p=0}^{k}\sum_{u=0}^{p} \frac{(-1)^{p+1}(2K)!}{4^{k}(k!)^2(2k-1)}\binom{k}{p}\binom{p}{u}(\beta_{12})^{-4p+2}(r_{2})^u(1 - (r_{2})^2)^{\frac{p - u}{2}}N^{12}_t(p, u). 
\end{eqnarray}
where
\begin{equation} 
   N^{12}_t(p, u) = E\bigg[\Big((\sigma_{1})^2_0 + \int_{0}^{\lambda T} e^{s}dZ^{1}_s \Big)^{p + u}\bigg] E\bigg[\Big(\frac{(\sigma_{2})^2_0 - r_{2}(\sigma_{2})^2_0}{\sqrt{1 - r_{2}}} + \int_{0}^{\lambda T} e^{s}dZ^{*}_s \Big)^{p - u}\bigg]. \nonumber
\end{equation}

$N^{12}_t(p, u)$ can be calculated using Theorem 1.
\item
Similarly $(\sigma_{1})^2_t$ and $(\sigma_{3})^2_t$ are defined respectively by \eqref{eq3} and $Z^{1}$, $Z^{3}$ and $Z^{**}$ are related by \eqref{eq5}. Let there exists $\beta_{31} >0$ such that for $0 \leq t \leq T$, $(\sigma_{1})^2_t, (\sigma_{3})^2_t \leq \beta^2_{31}$. Then,
\begin{eqnarray}\label{eq10}
    E[(\sigma_{1})_t(\sigma_{3})_t] = e^{-\lambda t}\sum_{k=0}^{\infty}\sum_{p=0}^{k}\sum_{u=0}^{p} \frac{(-1)^{p+1}(2K)!}{4^{k}(k!)^2(2k-1)}\binom{k}{p}\binom{p}{u}(\beta_{31})^{-4p+2}(r_{3})^u(1 - (r_{3})^2)^{\frac{p - u}{2}}N^{31}_t(p, u).  
\end{eqnarray}
where
\begin{equation}
   N^{31}_t(p, u) = E\bigg[\Big((\sigma_{1})^2_0 + \int_{0}^{\lambda T} e^{s}dZ^{1}_s \Big)^{p + u}\bigg] E\bigg[\Big(\frac{(\sigma_{3})^2_0 - r_{3}(\sigma_{1})^2_0}{\sqrt{1 - r_{3}}} + \int_{0}^{\lambda T} e^{s}dZ^{**}_s \Big)^{p - u}\bigg]. \nonumber
\end{equation}

$N^{31}_t(p, u)$ can be calculated using Theorem 1.
\end{enumerate}
\end{theorem}
Extending Theorem 3 we state the following,
\begin{theorem}
Suppose that $(\sigma_{2})^2_t$ and $(\sigma_{3})^2_t$ defined respectively by \eqref{eq3} and $Z^{2}$, $Z^{3}$, $Z^{*}$ and $Z^{**}$ are related by \eqref{eq6}. Let there exists $\beta_{23} >0$ such that for $0 \leq t \leq T$, $(\sigma_{2})^2_t, (\sigma_{3})^2_t \leq \beta^2_{12}$. Then, we can write,

\begin{eqnarray}\label{eq13}
  E[(\sigma_2)_t(\sigma_3)_t] &=& e^{-\lambda t} \sum_{k=0}^{\infty}\frac{(-1)^{p+1}(2K)!}{4^{k}(k!)^2(2k-1)} \sum_{p=0}^{k}\binom{k}{p}\sum_{u=0}^{p}\binom{p}{u}\sum_{v=0}^{u}\binom{u}{v} \sum_{w=0}^{u-v}\binom{u}{v}\Big(\frac{r_3}{r_2}\Big)^{u-w}\Big(1 - r^2_2\Big)^{\frac{p-v-w}{2}}\Big(1 - r^2_3\Big)^{\frac{p - u + w}{2}} \nonumber \\&& (\beta_{23})^{- 4p +2} N^{23}_{t}(p,u,v,w).  
\end{eqnarray}
where,
\begin{equation}
    N^{23}_{t}(p,u,v,w)  = E\Big(F^{u+v}_t\Big) E\Big((\Bar{G}^{**}_t )^{p - u + w}\Big) E\Big((G^{*}_t)^{p-v-w}\Big). \nonumber
\end{equation}
\end{theorem}

\textit{Proof: } It is clear from Equation \eqref{eq3} that,

\begin{equation}
    (\sigma_2)^{2}_t = e^{-\lambda t}(\sigma_2)^{2}_0 + e^{-\lambda t} \int_{0}^{ t} e^{\lambda s} d Z^{2}_{\lambda s}, \nonumber
\end{equation}
\begin{equation}
    (\sigma_2)^{2}_t = e^{-\lambda t}(\sigma_2)^{2}_0 + e^{-\lambda t} \int_{0}^{\lambda t} e^{s} \Big(r_{2}d Z^{1 }_{s} + \sqrt{1 - (r_{2})^2} dZ^{*}_s\Big), \nonumber
\end{equation}
\begin{equation}
    (\sigma_2)^{2}_t = e^{-\lambda t} \beta^2_{23}\bigg(\frac{(\sigma_2)^{2}_0 + r_{2}\int_{0}^{\lambda t} e^{s} d Z^{1}_{s} + \sqrt{1 - (r_{2})^2}\int_{0}^{\lambda t} e^{s} dZ^{*}_s}{\beta^2_{23}}\bigg). \nonumber
\end{equation}
Likewise,
\begin{equation}
    (\sigma_3)^{2}_t = e^{-\lambda t} \beta^2_{23}\bigg(\frac{(\sigma_3)^{2}_0 + r_{3}\int_{0}^{\lambda t} e^{s} d Z^{1}_{s} + \sqrt{1 - (r_{3})^2}\int_{0}^{\lambda t} e^{s} dZ^{**}_s}{\beta^2_{23}}\bigg). \nonumber
\end{equation}
We denote $F_t = (\sigma_2)^{2}_0 + r_{2}\int_{0}^{\lambda t} e^{s} d Z^{1}_{s}$, $G^{*}_t = \int_{0}^{\lambda t} e^{s} dZ^{*}_s$, $G^{**}_t = \int_{0}^{\lambda t} e^{s} dZ^{**}_s$ and $c = (\sigma_3)^{2}_0 - \frac{r_3}{r_2}(\sigma_2)^{2}_0$. \newline

Therefore, we can write,
\begin{equation}
    (\sigma_2)_t(\sigma_3)_t = e^{-\lambda t}\beta^2_{23}\sqrt{1 + \frac{F_t c + \frac{r_3}{r_2}F^2_t + \sqrt{1 - r^2_3}F_t G^{**}_t + \sqrt{1 - r^2_2}G^{*}_t c + \frac{r_3}{r_2}\sqrt{1 - r^2_2}G^{*}_t F_t +  \sqrt{1 - r^2_2} \sqrt{1 - r^2_3}G^{*}_t G^{**}_t - \beta^4_{23}}{\beta^4_{23}}}. \nonumber
\end{equation}

By Construction,
\begin{equation}
   \Bigg|\frac{F_t c + \frac{r_3}{r_2}F^2_t + \sqrt{1 - r^2_3}F_t G^{**}_t + \sqrt{1 - r^2_2}G^{*}_t c + \frac{r_3}{r_2}\sqrt{1 - r^2_2}G^{*}_t F_t +  \sqrt{1 - r^2_2} \sqrt{1 - r^2_3}G^{*}_t G^{**}_t - \beta^4_{23}}{\beta^4_{23}}\bigg| < 1. \nonumber
\end{equation}

Using the following convergence,

\begin{equation}
    \sqrt{1 + x} = \sum_{k=0}^{\infty}\frac{(-1)^{p+1}(2K)!}{4^{k}(k!)^2(2k-1)} x^k, \quad |x|<1 \nonumber
\end{equation}
We get,
\begin{eqnarray}
(\sigma_2)_t(\sigma_3)_t &=& e^{-\lambda t} \sum_{k=0}^{\infty}\frac{(-1)^{p+1}(2K)!}{4^{k}(k!)^2(2k-1)}\frac{1}{\beta^{4k-2}_{23}}  \bigg(F_t c + \frac{r_3}{r_2}F^2_t + \sqrt{1 - r^2_3}F_t G^{**}_t + \sqrt{1 - r^2_2}G^{*}_t c +\nonumber \\&& \frac{r_3}{r_2}\sqrt{1 - r^2_2}G^{*}_t F_t +  \sqrt{1 - r^2_2} \sqrt{1 - r^2_3}G^{*}_t G^{**}_t - \beta^4_{23}\bigg)^k, \nonumber 
\end{eqnarray}
\begin{eqnarray}
(\sigma_2)_t(\sigma_3)_t &=& e^{-\lambda t} \sum_{k=0}^{\infty}\frac{(-1)^{p+1}(2K)!}{4^{k}(k!)^2(2k-1)\beta^{4k-2}_{23}} \sum_{p=0}^{k}\binom{k}{p}\Big(F_t c + \frac{r_3}{r_2}F^2_t + \sqrt{1 - r^2_3}F_t G^{**}_t + \sqrt{1 - r^2_2}G^{*}_t c + \nonumber \\&& \frac{r_3}{r_2}\sqrt{1 - r^2_2}G^{*}_t F_t +  \sqrt{1 - r^2_2} \sqrt{1 - r^2_3}G^{*}_t G^{**}_t \Big)^p(- \beta_{23})^{4k - 4p}, \nonumber 
\end{eqnarray}
\begin{eqnarray}
&=& e^{-\lambda t} \sum_{k=0}^{\infty}\frac{(-1)^{p+1}(2K)!}{4^{k}(k!)^2(2k-1)} \sum_{p=0}^{k}\binom{k}{p}\sum_{u=0}^{p}\binom{k}{p}\Big(F_t c + \frac{r_3}{r_2}F^2_t + \sqrt{1 - r^2_3}F_t G^{**}_t + \frac{r_3}{r_2}\sqrt{1 - r^2_2}G^{*}_t F_t\Big)^u \nonumber \\&& \Big(\sqrt{1 - r^2_2}G^{*}_t c +   \sqrt{1 - r^2_2} \sqrt{1 - r^2_3}G^{*}_t G^{**}_t \Big)^{p - u}(\beta_{23})^{- 4p +2}, \nonumber 
\end{eqnarray}
\begin{eqnarray}
&=& e^{-\lambda t} \sum_{k=0}^{\infty}\frac{(-1)^{p+1}(2K)!}{4^{k}(k!)^2(2k-1)} \sum_{p=0}^{k}\binom{k}{p}\sum_{u=0}^{p}\binom{k}{p}\Big(F_t c + \frac{r_3}{r_2}F^2_t + \sqrt{1 - r^2_3}F_t G^{**}_t + \frac{r_3}{r_2}\sqrt{1 - r^2_2}G^{*}_t F_t\Big)^u \nonumber \\&& \Big(\sqrt{1 - r^2_2}G^{*}_t\Big)^{p-u}\Big( c + \sqrt{1 - r^2_3} G^{**}_t \Big)^{p - u}(\beta_{23})^{- 4p +2}, \nonumber 
\end{eqnarray}
\begin{eqnarray}
&=& e^{-\lambda t} \sum_{k=0}^{\infty}\frac{(-1)^{p+1}(2K)!}{4^{k}(k!)^2(2k-1)} \sum_{p=0}^{k}\binom{k}{p}\sum_{u=0}^{p}\binom{k}{p}F^u_t\Big(c + \frac{r_3}{r_2}F_t + \sqrt{1 - r^2_3}G^{**}_t + \frac{r_3}{r_2}\sqrt{1 - r^2_2}G^{*}_t\Big)^u \nonumber \\&& \Big(\sqrt{1 - r^2_2}G^{*}_t\Big)^{p-u}\Big( c + \sqrt{1 - r^2_3} G^{**}_t \Big)^{p - u}(\beta_{23})^{- 4p +2}, \nonumber 
\end{eqnarray}
\begin{eqnarray}
&=& e^{-\lambda t} \sum_{k=0}^{\infty}\frac{(-1)^{p+1}(2K)!}{4^{k}(k!)^2(2k-1)} \sum_{p=0}^{k}\binom{k}{p}\sum_{u=0}^{p}\binom{p}{u}F^u_t\sum_{v=0}^{u}\binom{u}{v}\Big(\frac{r_3}{r_2}F_t\Big)^v \Big(c + \sqrt{1 - r^2_3}G^{**}_t + \frac{r_3}{r_2}\sqrt{1 - r^2_2}G^{*}_t\Big)^{u-v} \nonumber \\&&
\Big(\sqrt{1 - r^2_2}G^{*}_t\Big)^{p-u}\Big( c + \sqrt{1 - r^2_3} G^{**}_t \Big)^{p - u}(\beta_{23})^{- 4p +2}, \nonumber
\end{eqnarray}
\begin{eqnarray}
&=& e^{-\lambda t} \sum_{k=0}^{\infty}\frac{(-1)^{p+1}(2K)!}{4^{k}(k!)^2(2k-1)} \sum_{p=0}^{k}\binom{k}{p}\sum_{u=0}^{p}\binom{p}{u}F^u_t\sum_{v=0}^{u}\binom{u}{v}\Big(\frac{r_3}{r_2}F_t\Big)^v \sum_{w=0}^{u-v}\binom{u}{v}\Big(c + \sqrt{1 - r^2_3}G^{**}_t \Big)^{w}\nonumber \\&& \Big(\frac{r_3}{r_2}\sqrt{1 - r^2_2}G^{*}_t\Big)^{u-v-w} \Big(\sqrt{1 - r^2_2}G^{*}_t\Big)^{p-u}\Big( c + \sqrt{1 - r^2_3} G^{**}_t \Big)^{p - u}(\beta_{23})^{- 4p +2}, \nonumber
\end{eqnarray}
\begin{eqnarray}\label{eq12}
&=& e^{-\lambda t} \sum_{k=0}^{\infty}\frac{(-1)^{p+1}(2K)!}{4^{k}(k!)^2(2k-1)} \sum_{p=0}^{k}\binom{k}{p}\sum_{u=0}^{p}\binom{p}{u}\sum_{v=0}^{u}\binom{u}{v} \sum_{w=0}^{u-v}\binom{u}{v}F^{u+v}_t\Big(\frac{r_3}{r_2}\Big)^{u-w}\Big(\Bar{G}^{**}_t \Big)^{p - u + w} \nonumber \\&& \Big(1 - r^2_2\Big)^{\frac{p-v-w}{2}}\Big(1 - r^2_3\Big)^{\frac{p -u + w}{2}} 
  \Big(G^{*}_t\Big)^{p-v-w} (\beta_{23})^{- 4p +2}.
\end{eqnarray}
where 
\begin{equation}
\Bar{G}^{**}_t = G^{**}_t + \frac{c}{\sqrt{1-r^2_3}}. \nonumber    
\end{equation}
Taking the expectation of Equation \eqref{eq12}
\begin{eqnarray}
  E[(\sigma_2)_t(\sigma_3)_t] &=& e^{-\lambda t} \sum_{k=0}^{\infty}\frac{(-1)^{p+1}(2K)!}{4^{k}(k!)^2(2k-1)} \sum_{p=0}^{k}\binom{k}{p}\sum_{u=0}^{p}\binom{p}{u}\sum_{v=0}^{u}\binom{u}{v} \sum_{w=0}^{u-v}\binom{u}{v}\Big(\frac{r_3}{r_2}\Big)^{u-w}\Big(1 - r^2_2\Big)^{\frac{p-v-w}{2}}\nonumber \\&& \Big(1 - r^2_3\Big)^{\frac{p - u + w}{2}}  (\beta_{23})^{- 4p +2} E\Big(F^{u+v}_t\Big) E\Big((\Bar{G}^{**}_t )^{p - u + w}\Big) E\Big((G^{*}_t)^{p-v-w}\Big). \nonumber
\end{eqnarray}
 
Assigning,
 
\begin{equation}
    N^{23}_{t}(p,u,v,w)  = E\Big(F^{u+v}_t\Big) E\Big((\Bar{G}^{**}_t )^{p - u + w}\Big) E\Big((G^{*}_t)^{p-v-w}\Big). \nonumber
\end{equation}

$N^{23}_{t}(p,u,v,w)$ can be computed by using Theorem 1. Comparing the Equation \eqref{eq15} with different terms of $N^{23}_{t}(p,u,v,w)$, we get,\newline
For $F_t$,
\begin{equation}
     \alpha_1 = \frac{(\sigma_2)^{2}_0}{r_2}. \nonumber
\end{equation}
\begin{equation}
     V_t = Z^{1}_t \nonumber
\end{equation} 
For $G'^{**}_t$,
\begin{equation}
\alpha_1 = \frac{(\sigma_3)^{2}_0 - \frac{r_3}{r_2}(\sigma_2)^{2}_0}{\sqrt{1-r^2_3}}. \nonumber 
\end{equation}
\begin{equation}
    V_t = Z^{**}_t. \nonumber
\end{equation}
For $G'^{*}_t$,
\begin{equation}
\alpha_1 = 0. \nonumber 
\end{equation}
\begin{equation}
   V_t = Z^{*}_t. \nonumber
\end{equation}

Thus, we can write $N^{23}_{t}(p,u,v,w)$ as
\begin{equation} \label{eq16}
    N^{23}_{t}(p,u,v,w)  = r_2^{u+v}E\bigg[\Big(\frac{(\sigma_2)^{2}_0}{r_2} + \int_{0}^{\lambda T} e^{s}d Z^{1}_s\Big)^{u+v}\bigg] E\bigg[\Big(\frac{(\sigma_3)^{2}_0 - \frac{r_3}{r_2}(\sigma_2)^{2}_0}{\sqrt{1-r^2_3}} + \int_{0}^{\lambda T} e^{s}d Z^{**}_s\Big)^{p - u + w}\bigg] E\bigg[\Big(\int_{0}^{\lambda T} e^{s}d Z^{*}_s\Big)^{p-v-w}\bigg]. 
\end{equation} 

which concludes the proof of Theorem 4.$\qed$ \newline

Now considering $[X^{1}_T, X^{2}_T]$ of Equation \eqref{eq7} and substituting from Equation \eqref{eq8} we get,

\begin{equation}
    E [Cov(S^{1},S^{2})] = \frac{1}{T}E[X^{1}_T, X^{2}_T] = \frac{1}{T}\int_{0}^{T} \gamma_{12} E\Big((\sigma_{1})_t (\sigma_{2})_t \Big) dt +  \frac{1}{T}\rho_{1}\rho_{2}E \Big(\int_{0}^{\lambda T} \int_{0}^{\infty} y^2 J_{Z^1}(ds, dy)\Big). \nonumber
\end{equation}

\begin{equation}
   \frac{1}{T}\int_{0}^{T} \gamma_{12} E\Big((\sigma_{1})_t (\sigma_{2})_t \Big) dt = \frac{\gamma_{12}}{T} \int_{0}^{T} e^{-\lambda t}\sum_{k=0}^{\infty}\sum_{p=0}^{k}\sum_{u=0}^{p} \frac{(-1)^{p+1}(2K)!}{4^{k}(k!)^2(2k-1)}\binom{k}{p}\binom{p}{u}(\beta_{12})^{-4p+2}(r_{2})^u(1 - (r_{2})^2)^{\frac{p - u}{2}}N^{12}_t(p, u) dt, \nonumber
\end{equation}
\begin{equation}
   = \frac{\gamma_{12}}{T} \sum_{k=0}^{\infty}\sum_{p=0}^{k}\sum_{u=0}^{p} \frac{(-1)^{p+1}(2K)!}{4^{k}(k!)^2(2k-1)}\binom{k}{p}\binom{p}{u}(\beta_{12})^{-4p+2}(r_{2})^u(1 - (r_{2})^2)^{\frac{p - u}{2}}\int_{0}^{T} e^{-\lambda t} N^{12}_t(p, u) dt. \nonumber
\end{equation}
Observing $E(J_{Z^1}(ds, dy)) = \nu_{Z^1}(dy) ds$ we obtain,
\begin{equation}
   \frac{\rho_{1}\rho_{2}}{T}E \Big(\int_{0}^{\lambda T} \int_{0}^{\infty} y^2 J_{Z^1}(ds, dy)\Big) = \rho_{1}\rho_{2}\lambda \int_{0}^{\infty} y^2 \nu_Z^1(dy). \nonumber
\end{equation}
Now from the paper of \citeauthor{Cont}(\citeyear{Cont}) (proposition 3.13) we can see $\kappa^{1}_2 = Var(Z^{1}) = \int_{0}^{\infty}y^2 \nu_{Z^1}(dy)$.
\begin{equation}
   \frac{\rho_{1}\rho_{2}}{T}E \Big(\int_{0}^{\lambda T} \int_{0}^{\infty} y^2 J_{Z^1}(ds, dy)\Big) = \rho_{1}\rho_{2}\lambda \kappa^{1}_2. \nonumber 
\end{equation}
Therefore,
\begin{eqnarray}\label{eq17}
    E [Cov(S^{1},S^{2})] &=& \frac{\gamma_{12}}{T} \sum_{k=0}^{\infty}\sum_{p=0}^{k}\sum_{u=0}^{p} \frac{(-1)^{p+1}(2K)!}{4^{k}(k!)^2(2k-1)}\binom{k}{p}\binom{p}{u}(\beta_{12})^{-4p+2}(r_{2})^u(1 - (r_{2})^2)^{\frac{p - u}{2}} \nonumber \\&& \int_{0}^{T} e^{-\lambda t} N^{12}_t(p, u) dt  + \rho_{1}\rho_{2}\lambda \kappa^{1}_2 = E [Cov(S^{2},S^{1})]. 
\end{eqnarray}
Similarly using Equations \eqref{eq7} and \eqref{eq13},
\begin{eqnarray}\label{eq18}
    E [Cov(S^{2},S^{3})] &=& \frac{\gamma_{23}}{T} \sum_{k=0}^{\infty}\frac{(-1)^{p+1}(2K)!}{4^{k}(k!)^2(2k-1)} \sum_{p=0}^{k}\binom{k}{p}\sum_{u=0}^{p}\binom{p}{u}\sum_{v=0}^{u}\binom{u}{v} \sum_{w=0}^{u-v}\binom{u}{v}\Big(\frac{r_3}{r_2}\Big)^{u-w}\Big(1 - r^2_2\Big)^{\frac{p-v-w}{2}}\Big(1 - r^2_3\Big)^{\frac{p - u + w}{2}}\nonumber \\&& (\beta_{23})^{- 4p +2}  \int_{0}^{T} e^{-\lambda t} N^{23}_t(p, u, v, w) dt  + \rho_{2}\rho_{3}\lambda \kappa^{1}_2 = E [Cov(S^{3},S^{2})].
\end{eqnarray}
And again using Equations \eqref{eq7} and \eqref{eq10},
\begin{eqnarray}\label{eq19}
    E [Cov(S^{3},S^{1})] &=& \frac{\gamma_{31}}{T} \sum_{k=0}^{\infty}\sum_{p=0}^{k}\sum_{u=0}^{p} \frac{(-1)^{p+1}(2K)!}{4^{k}(k!)^2(2k-1)}\binom{k}{p}\binom{p}{u}(\beta_{31})^{-4p+2}(r_{3})^u(1 - (r_{3})^2)^{\frac{p - u}{2}} \nonumber \\&& \int_{0}^{T} e^{-\lambda t} N^{31}_t(p, u) dt  + \rho_{3}\rho_{1}\lambda \kappa^{1}_2 = E [Cov(S^{1},S^{3})]. 
\end{eqnarray}

Now we are going to calculate,

\begin{equation}
    [X^{1}_T, X^{1}_T] = \int_{0}^{T}  (\sigma_{1})^2_t dt +  (\rho_{1})^2\int_{0}^{\lambda T} \int_{0}^{\infty} y^2 J_{Z^1}(ds, dy). \nonumber
\end{equation}
Similarly,
\begin{equation}
    [X^{2}_T, X^{2}_T] = \int_{0}^{T}  (\sigma_{2})^2_t dt +  (\rho_{2})^2\int_{0}^{\lambda T} \int_{0}^{\infty} y^2 J_{Z^1}(ds, dy). \nonumber
\end{equation}
\begin{equation}\label{eq20}
    [X^{3}_T, X^{3}_T] = \int_{0}^{T}  (\sigma_{3})^2_t dt +  (\rho_{3})^2\int_{0}^{\lambda T} \int_{0}^{\infty} y^2 J_{Z^1}(ds, dy). 
\end{equation}

We will now try to find the value for 
\begin{equation}
 E [Cov(S^{1},S^{1})] =  E [Var (S^{1})] = \frac{1}{T}E[X^{1}_T, X^{1}_T]  \nonumber 
\end{equation}
\begin{equation}
\frac{1}{T}E[X^{1}_T, X^{1}_T] = \frac{1}{T}\int_{0}^{T}  E[(\sigma_{1})^2_t] dt + \frac{1}{T}(\rho_{1})^2 E \Big[\int_{0}^{\lambda T}\int_{0}^{\infty} y^2 J_{Z^1}(ds, dy)\Big]. \nonumber 
\end{equation}

Using the solution of \eqref{eq3} 

\begin{equation}
    (\sigma_1)^{2}_t = e^{-\lambda t}(\sigma_1)^{2}_0 + e^{-\lambda t} \int_{0}^{ t} e^{\lambda s} d Z^{1}_{\lambda s}. \nonumber
\end{equation}
\begin{equation}
    E[(\sigma_1)^{2}_t] = e^{-\lambda t}E\big[(\sigma_1)^{2}_0 + \int_{0}^{ t} e^{\lambda s} d Z^{1}_{\lambda s}\big]. \nonumber
\end{equation}
Comparing this with \eqref{eq15}, we have $\alpha_1 = (\sigma_1)^{2}_0$ and $V_t = Z^{1}_t$ so,
\begin{equation}
    N^{1}_t = E\bigg[(\sigma_1)^{2}_0 + \int_{0}^{ t} e^{\lambda s} d Z^{1}_{\lambda s}\bigg].  \nonumber
\end{equation}
\begin{equation}
    \frac{1}{T}\int_{0}^{T}  E[(\sigma_{1})^2_t] dt = \frac{1}{T}\int_{0}^{T}e^{-\lambda t} N^{1}_t dt. \nonumber
\end{equation}
\begin{equation}
    \frac{1}{T}(\rho_{1})^2 E \Big[\int_{0}^{\lambda T}\int_{0}^{\infty} y^2 J_{Z^1}(ds, dy)\Big] = \frac{(\rho_{1})^2}{T}\lambda \kappa^{1}_2. \nonumber
\end{equation}

Therefore the value obtained is,

\begin{equation}
E [Var (S^{1})] = \frac{1}{T}E[X^{1}_T, X^{1}_T] = \frac{1}{T}\int_{0}^{T}e^{-\lambda t} N^{1}_t dt + \frac{(\rho_{1})^2}{T}\lambda \kappa^{1}_2. \nonumber 
\end{equation}

Similarly,

\begin{equation}
 E [Cov(S^{2},S^{2})] = E [Var (S^{2})] = \frac{1}{T}E[X^{2}_T, X^{2}_T] = \frac{1}{T}\int_{0}^{T}e^{-\lambda t} N^{2}_t dt + \frac{(\rho_{2})^2}{T}\lambda \kappa^{1}_2. \nonumber 
\end{equation}

where 
\begin{equation}
    N^{2}_t = E\bigg[(\sigma^2)^{2}_0 + \int_{0}^{ t} e^{\lambda s} d Z^{1}_{\lambda s}\bigg].  \nonumber
\end{equation}

\begin{equation}
 E [Cov(S^{3},S^{3})] = E [Var (S^{3})] = \frac{1}{T}E[X^{3}_T, X^{3}_T] = \frac{1}{T}\int_{0}^{T}e^{-\lambda t} N^{3}_t dt + \frac{(\rho_{3})^2}{T}\lambda \kappa^{1}_2. \nonumber 
\end{equation}

where
\begin{equation}
    N^{3}_t = E\bigg[(\sigma^3)^{2}_0 + \int_{0}^{ t} e^{\lambda s} d Z^{1}_{\lambda s}\bigg].  \nonumber
\end{equation}

To avoid complexity of the numerical computations we derive an alternative approximate version of Equations \eqref{eq17}, \eqref{eq18} and \eqref{eq19}. 
\begin{Results}
A useful estimate approximation regarding the expected value of realized volatility $(\sigma_R)$ is obtained in \citeauthor{Ref2}(\citeyear{Ref2}) and is given by
\begin{equation}\label{eq29}
    E[\sigma_R] = E[\sqrt{\sigma^2_R}]\approx \sqrt{E[\sigma^2_R]} - \frac{Var[\sigma^2_R]}{8(E[\sigma^2_R])^{\frac{3}{2}}}.
\end{equation}
\end{Results}

\begin{theorem}(Lemma 3.9 from \citeauthor{sengupta}(\citeyear{sengupta}))
\begin{enumerate}
    \item 
For $Z^{1}_t, Z^{2}_t$ and $Z^{*}_t$ related by \eqref{eq4} we can calculate ,
\begin{eqnarray}\label{eq25}
 E\Big[(\sigma_1)^2_t(\sigma_2)^2_t\Big] &=& e^{-2\lambda t}(\sigma_1)^2_0(\sigma_2)^2_0 + e^{-2\lambda t}\Big((\sigma_1)^2_0 \sqrt{1 - (r_{2})^2}\kappa^{*}_1\Big)  + \Big((\sigma_1)^2_0 r_2 + (\sigma_2)^2_0 ) \kappa^{1}_1 \Big) \Big(e^{\lambda t} - 1\Big) + \nonumber \\&& 
 e^{-2\lambda t}\bigg(r_{2}\Big((\kappa^1_1)^2(e^{\lambda t} - 1)^2 + \frac{\kappa^1_2}{2}(e^{2\lambda t} - 1)\Big) + \sqrt{1 - (r_{2})^2}\kappa^1_1\kappa^{*}_1(e^{\lambda t} - 1)^2 \bigg).
\end{eqnarray}
\begin{eqnarray}
 Var\Big[(\sigma_1)^2_t(\sigma_2)^2_t\Big] &=& Var\Big((\sigma_1)^2_0(\sigma_2)^2_0 + ((\sigma_1)^2_0\sqrt{1 - r^2_2})Y_2 + ((\sigma_2)^2_0 + r_2(\sigma_1)^2_0) Y_1 + r_2 Y^2_1 + \sqrt{1 - r^2_2}Y_1Y_2\Big). \nonumber
\end{eqnarray}
\begin{eqnarray}\label{eq26}
 Var\Big[(\sigma_1)^2_t(\sigma_2)^2_t\Big] &=&  \Big((\sigma_1)^2_0\sqrt{1 - r^2_2}\Big)^2 Var(Y_2) + \Big((\sigma_2)^2_0 + r_2(\sigma_1)^2_0\Big)^2 Var(Y_1) + r^2_2 Var(Y^2_1) + (1 - r^2_2)Var(Y_1Y_2) + \nonumber \\&& 2\Big((\sigma_1)^2_0\sqrt{1 - r^2_2}\Big)\Big(\sqrt{1 - r^2_2}\Big) Cov(Y_2,Y_1Y_2) + 2\Big((\sigma_2)^2_0 + r_2(\sigma_1)^2_0\Big)r_2 Cov(Y_1,Y^2_1) + \nonumber \\&& 2\Big((\sigma_2)^2_0 + r_2(\sigma_1)^2_0\Big)\Big(\sqrt{1 - r^2_2}\Big)Cov(Y_1, Y_1Y_2) + 2r_2\Big(\sqrt{1 - r^2_2}\Big)Cov(Y^2_1,Y_1Y_2). 
\end{eqnarray}
\item
Similarly when $Z^1_t, Z^3_t$ and $Z^{**}_t$ are related as \eqref{eq5} we can calculate,
\begin{eqnarray}\label{eq27}
 E\Big[(\sigma_1)^2_t(\sigma_3)^2_t\Big] &=& e^{-2\lambda t}(\sigma_1)^2_0(\sigma_3)^2_0 + e^{-2\lambda t}\Big((\sigma_1)^2_0 \sqrt{1 - (r_{3})^2}\kappa^{**}_1\Big)  + \Big((\sigma_1)^2_0 r_3 + (\sigma_3)^2_0 ) \kappa^{1}_1 \Big) \Big(e^{\lambda t} - 1\Big) + \nonumber \\&& 
 e^{-2\lambda t}\bigg(r_{3}\Big((\kappa^1_1)^2(e^{\lambda t} - 1)^2 + \frac{\kappa^1_2}{2}(e^{2\lambda t} - 1)\Big) + \sqrt{1 - (r_{3})^2}\kappa^1_1\kappa^{**}_1(e^{\lambda t} - 1)^2 \bigg).
\end{eqnarray}
\begin{eqnarray}
 Var\Big[(\sigma_1)^2_t(\sigma_3)^2_t\Big] &=& Var\Big((\sigma_1)^2_0(\sigma_3)^2_0 + ((\sigma_1)^2_0\sqrt{1 - r^2_3})Y_3 + ((\sigma_3)^2_0 + r_3(\sigma_1)^2_0) Y_1 + r_3 Y^2_1 + \sqrt{1 - r^2_3}Y_1Y_3\Big), \nonumber
\end{eqnarray}
\begin{eqnarray}\label{eq28}
 Var\Big[(\sigma_1)^2_t(\sigma_3)^2_t\Big] &=&  \Big((\sigma_1)^2_0\sqrt{1 - r^2_3}\Big)^2 Var(Y_3) + \Big((\sigma_3)^2_0 + r_3(\sigma_1)^2_0\Big)^2 Var(Y_1) + r^2_3 Var(Y^2_1) + (1 - r^2_3)Var(Y_1Y_3) + \nonumber \\&& 2\Big((\sigma_1)^2_0\sqrt{1 - r^2_3}\Big)\Big(\sqrt{1 - r^2_3}\Big) Cov(Y_3,Y_1Y_3) + 2\Big((\sigma_3)^2_0 + r_3(\sigma_1)^2_0\Big)r_3 Cov(Y_1,Y^2_1) + \nonumber \\&& 2\Big((\sigma_3)^2_0 + r_3(\sigma_1)^2_0\Big)\Big(\sqrt{1 - r^2_3}\Big)Cov(Y_1, Y_1Y_3) + 2r_3\Big(\sqrt{1 - r^2_3}\Big)Cov(Y^2_1,Y_1Y_3). 
\end{eqnarray}

\end{enumerate}
\end{theorem}
Extending Theorem 5 we state a theorem pertaining to our set up..
\begin{theorem}
For $Z^{2}_t, Z^{3}_t, Z^{*}_t$ and $Z^{**}_t$ related by \eqref{eq6} we get,

\begin{eqnarray}\label{eq22}
 E\Big[(\sigma_2)^2_t(\sigma_3)^2_t\Big] &=& e^{-2\lambda t}(\sigma_2)^2_0(\sigma_3)^2_0 + e^{-2\lambda t}\Big((\sigma_2)^2_0\Big(r_3\kappa^{1}_1 + \sqrt{1 - (r_{3})^2}\kappa^{**}_1\Big)\Big(e^{\lambda t} - 1\Big)  + (\sigma_3)^2_0\Big(r_2\kappa^{1}_1 + \sqrt{1 - (r_{2})^2}\kappa^{*}_1\Big)\nonumber \\&& \Big(e^{\lambda t} - 1\Big) + e^{-2\lambda t}r_{2}r_{3}\Big((\kappa^1_1)^2(e^{\lambda t} - 1)^2 + \frac{\kappa^1_2}{2}(e^{2\lambda t} - 1)\Big) + r_{2}\sqrt{1 - (r_{3})^2}\kappa^1_1\kappa^{**}_1(e^{\lambda t} - 1)^2 + \nonumber \\&& r_{3}\sqrt{1 - (r_{2})^2} \kappa^1_1\kappa^*_1(e^{-\lambda t} - 1)^2 + \sqrt{1 - (r_{2})^2}\sqrt{1 - (r_{3})^2} \kappa^{**}_1\kappa^*_1(e^{\lambda t} - 1)^2.
\end{eqnarray}
\begin{eqnarray}\label{eq24}
    Var[(\sigma_2)^2_t(\sigma_3)^2_t] &=&  a^2_2Var(Y_1) + a^2_3Var(Y_3) + a^2_4var(Y_2) + a^2_5Var(Y^2_1) + a^2_6Var(Y_1Y_3) + a^2_7Var(Y_1Y_2) +  2a_2a_5Cov(Y_1,Y^2_1) + \nonumber \\&&  2a_2a_6Cov(Y_1,Y_1Y_3) + 2a_2a_7Cov(Y_1, Y_1Y_2)  + 2a_3a_6Cov(Y_3,Y_1Y_3) +   2a_4a_7Cov(Y_2,Y_1Y_2) \nonumber \\&& + 2a_5a_6Cov(Y^2_1,Y_1Y_3)  + 2a_5a_7Cov(Y^2_1,Y_1Y_2) + 2a_6a_7Cov(Y_1Y_3,Y_1Y_2).
\end{eqnarray}
\end{theorem}

\textit{Proof: }To start with we are going to find the product of the pairs of variances.\newline

Calculating the product of $(\sigma_2)^2_t$ and $(\sigma_3)^2_t$ defined in Equation \eqref{eq3}
\begin{eqnarray}\label{eq21}
 (\sigma_2)^2_t(\sigma_3)^2_t &=& e^{-2\lambda t}(\sigma_2)^2_0(\sigma_3)^2_0 + e^{-2\lambda t}\Big((\sigma_2)^2_0\int_{0}^{\lambda t}e^{s}dZ^3_{s} + (\sigma_3)^2_0\int_{0}^{\lambda t}e^{s}dZ^2_{s}\Big) + e^{-2\lambda t}\int_{0}^{\lambda t}e^{s}dZ^2_{s}\int_{0}^{\lambda t}e^{s}dZ^3_{s}. \nonumber \\&&
\end{eqnarray}
\begin{equation}
    E\Big(\int_{0}^{\lambda t}e^{s}dZ^2_{s}\Big) = E\Big(\int_{0}^{\lambda t}e^{s}(r_{2}d Z^{1 }_{s} + \sqrt{1 - (r_{2})^2} dZ^{*}_s)\Big) = \Big(r_2\kappa^{1}_1 + \sqrt{1 - (r_{2})^2}\kappa^{*}_1\Big)\Big(e^{-\lambda t} - 1\Big). \nonumber
\end{equation}
\begin{equation}
    E\Big(\int_{0}^{\lambda t}e^{s}dZ^3_{s}\Big) = E\Big(\int_{0}^{\lambda t}e^{s}(r_{3}d Z^{1 }_{s} + \sqrt{1 - (r_{3})^2} dZ^{**}_s)\Big) = \Big(r_3\kappa^{1}_1 + \sqrt{1 - (r_{3})^2}\kappa^{**}_1\Big)\Big(e^{-\lambda t} - 1\Big). \nonumber
\end{equation}
\begin{equation}
    E\Big(\int_{0}^{\lambda t}e^{s}dZ^2_{s}\int_{0}^{\lambda t}e^{s}dZ^3_{s}\Big) = E\Big(\int_{0}^{\lambda t}e^{s}(r_{2}d Z^{1 }_{s} + \sqrt{1 - (r_{2})^2} dZ^{*}_s)\int_{0}^{\lambda t}e^{s}(r_{3}d Z^{1 }_{s} + \sqrt{1 - (r_{3})^2} dZ^{**}_s)\Big).\nonumber
\end{equation}
\begin{equation}
    = r_{2}r_{3}E\Big(\int_{0}^{\lambda t}e^{s}d Z^{1}_{s}\Big)^2 + r_{2}\sqrt{1 - (r_{3})^2}E\Big(\int_{0}^{\lambda t}e^{s}d Z^{1}_{s}\Big)E\Big(\int_{0}^{\lambda t}e^{s}d Z^{**}_{s}\Big) + r_{3}\sqrt{1 - (r_{2})^2} E\Big(\int_{0}^{\lambda t}e^{s}d Z^{1}_{s}\Big)E\Big(\int_{0}^{\lambda t}e^{s}d Z^{*}_{s}\Big) \nonumber
\end{equation}
\begin{equation}
+ \sqrt{1 - (r_{2})^2}\sqrt{1 - (r_{3})^2} E\Big(\int_{0}^{\lambda t}e^{s}d Z^{*}_{s}\Big)E\Big(\int_{0}^{\lambda t}e^{s}d Z^{**}_{s}\Big),\nonumber
\end{equation}
\begin{equation}
    = r_{2}r_{3}\Big((\kappa^1_1)^2(e^{\lambda t} - 1)^2 + \frac{\kappa^1_2}{2}(e^{2\lambda t} - 1)\Big) + r_{2}\sqrt{1 - (r_{3})^2}\kappa^1_1\kappa^{**}_1(e^{\lambda t} - 1)^2 + r_{3}\sqrt{1 - (r_{2})^2} \kappa^1_1\kappa^*_1(e^{\lambda t} - 1)^2 \nonumber
\end{equation}
\begin{equation}
+ \sqrt{1 - (r_{2})^2}\sqrt{1 - (r_{3})^2} \kappa^{**}_1\kappa^*_1(e^{\lambda t} - 1)^2.\nonumber
\end{equation}

\begin{eqnarray}
 E\Big[(\sigma_2)^2_t(\sigma_3)^2_t\Big] &=& e^{-2\lambda t}(\sigma_2)^2_0(\sigma_3)^2_0 + e^{-2\lambda t}\Big((\sigma_2)^2_0\Big(r_3\kappa^{1}_1 + \sqrt{1 - (r_{3})^2}\kappa^{**}_1\Big)\Big(e^{\lambda t} - 1\Big)  + (\sigma_3)^2_0\Big(r_2\kappa^{1}_1 + \sqrt{1 - (r_{2})^2}\kappa^{*}_1\Big)\nonumber \\&& \Big(e^{\lambda t} - 1\Big) + e^{-2\lambda t}r_{2}r_{3}\Big((\kappa^1_1)^2(e^{\lambda t} - 1)^2 + \frac{\kappa^1_2}{2}(e^{2\lambda t} - 1)\Big) + r_{2}\sqrt{1 - (r_{3})^2}\kappa^1_1\kappa^{**}_1(e^{\lambda t} - 1)^2 + \nonumber \\&& r_{3}\sqrt{1 - (r_{2})^2} \kappa^1_1\kappa^*_1(e^{-\lambda t} - 1)^2 + \sqrt{1 - (r_{2})^2}\sqrt{1 - (r_{3})^2} \kappa^{**}_1\kappa^*_1(e^{\lambda t} - 1)^2.  \nonumber
\end{eqnarray}

Expanding the dependency of $d Z^{2}_s$ and $d Z^{3}_s$ in terms of the independent process in the Equation \eqref{eq21}

\begin{eqnarray}\label{eq23}
 (\sigma_2)^2_t(\sigma_3)^2_t &=& e^{-2\lambda t}(\sigma_2)^2_0(\sigma_3)^2_0 + e^{-2\lambda t}\Big((\sigma_2)^2_0 r_3\int_{0}^{\lambda t}e^{s}dZ^1_{s} + (\sigma_2)^2_0 \sqrt{1 - r^2_3}\int_{0}^{\lambda t}e^{s}dZ^{
 **}_{s} + (\sigma_3)^2_0r_2\int_{0}^{\lambda t}e^{s}dZ^1_{s} + \nonumber \\&& (\sigma_3)^2_0 \sqrt{1 - r^2_2}\int_{0}^{\lambda t}e^{s}dZ^{
 *}_{s}\Big) + e^{-2\lambda t}\bigg(r_{2}r_{3}\Big(\int_{0}^{\lambda t}e^{s}d Z^{1}_{s}\Big)^2 + r_{2}\sqrt{1 - (r_{3})^2}\Big(\int_{0}^{\lambda t}e^{s}d Z^{1}_{s}\Big)\Big(\int_{0}^{\lambda t}e^{s}d Z^{**}_{s}\Big) + \nonumber \\&& r_{3}\sqrt{1 - (r_{2})^2} \Big(\int_{0}^{\lambda t}e^{s}d Z^{1}_{s}\Big)\Big(\int_{0}^{\lambda t}e^{s}d Z^{*}_{s}\Big)\bigg).
\end{eqnarray}
Let us define
\begin{equation}
Y_1 = \int_{0}^{\lambda t}e^{s}dZ^{1}_{s}. \nonumber
\end{equation}
\begin{equation}
Y_2 = \int_{0}^{\lambda t}e^{s}dZ^{*}_{s}. \nonumber
\end{equation}
\begin{equation}
Y_3 = \int_{0}^{\lambda t}e^{s}dZ^{**}_{s}. \nonumber
\end{equation}
\begin{equation}
a_1 = (\sigma_2)^2_0(\sigma_3)^2_0. \nonumber
\end{equation}
\begin{equation}
a_2 = (\sigma_2)^2_0 r_3 + (\sigma_3)^2_0 r_2. \nonumber
\end{equation}
\begin{equation}
a_3 = (\sigma_2)^2_0 \sqrt{1 - r^2_3}. \nonumber
\end{equation}
\begin{equation}
a_4 = (\sigma_3)^2_0 \sqrt{1 - r^2_2}. \nonumber
\end{equation}
\begin{equation}
a_5 = r_2r_3. \nonumber
\end{equation}
\begin{equation}
a_6 = r_2 \sqrt{1 - r^2_3}. \nonumber
\end{equation}
\begin{equation}
a_7 = r_3\sqrt{1 - r^2_2}. \nonumber
\end{equation}
So, Variance of Equation \eqref{eq23} looks like
\begin{equation}
    Var[(\sigma_2)^2_t(\sigma_3)^2_t] = Var[a_1 + a_2Y_1 + a_3Y_3 + a_4Y_2 + a_5Y^2_1 + a_6Y_1Y_3 + a_7Y_1Y_2]. \nonumber
\end{equation}

Using the independence of $Y_1, Y_2$ and $Y_3$ and comparring with \eqref{eq15}, it is easy to show
\begin{equation}
    Var(Y^2_1) = E(Y^4_1) - (E(Y^2_1))^2. \nonumber
\end{equation}
\begin{equation}
    Var(Y_2) = E(Y^2_2) - (E(Y_2))^2. \nonumber
\end{equation}
\begin{equation}
    Var(Y_3) = E(Y^2_3) - (E(Y_3))^2. \nonumber
\end{equation}
\begin{eqnarray}
    E(Y^4_1) &=& (\kappa^{1}_1)^4(e^{\lambda t} - 1)^4 + 3(\kappa^1_1)^2\kappa^1_2(e^{\lambda t} - 1)^2(e^{2\lambda t} - 1) + \frac{3}{4}(\kappa^1_2)^2(e^{2\lambda t} - 1)^2 + \nonumber \\&& \frac{4}{3}(\kappa^1_3)(\kappa^1_4)(e^{\lambda t} - 1)(e^{3\lambda t} - 1) + \frac{\kappa^1_4}{4}(e^{4\lambda t} - 1). \nonumber
\end{eqnarray}
\begin{equation}
    E(Y^3_1) = (\kappa^{1}_1)^3(e^{\lambda t} - 1)^3 + \frac{3}{2}(\kappa^1_1)(\kappa^1_2)(e^{\lambda t} - 1)(e^{2\lambda t} - 1) + \frac{\kappa^1_3}{3}(e^{3\lambda t} - 1). \nonumber
\end{equation}
\begin{equation}
    E(Y^2_1) = (\kappa^{1}_1)^2(e^{\lambda t} - 1)^2 + \frac{\kappa^1_2}{2}(e^{2\lambda t} - 1). \nonumber
\end{equation}
\begin{equation}
    E(Y^2_2) = (\kappa^{*}_1)^2(e^{\lambda t} - 1)^2 + \frac{\kappa^*_2}{2}(e^{2\lambda t} - 1). \nonumber
\end{equation}
\begin{equation}
    E(Y^3_2) = (\kappa^{**}_1)^2(e^{\lambda t} - 1)^2 + \frac{\kappa^{**}_2}{2}(e^{2\lambda t} - 1). \nonumber
\end{equation}
\begin{equation}
    E(Y_1) = \kappa^{1}_1(e^{\lambda t} - 1). \nonumber
\end{equation}
\begin{equation}
    E(Y_2) = \kappa^{*}_1(e^{\lambda t} - 1). \nonumber
\end{equation}
\begin{equation}
    E(Y_3) = \kappa^{**}_1(e^{\lambda t} - 1). \nonumber
\end{equation}
\begin{equation}
    Var(Y_1Y_2) = E(Y^2_1)E(Y^2_2) - (E(Y_1))^2(E(Y_2))^2. \nonumber
\end{equation}
\begin{equation}
    Var(Y_1Y_3) = E(Y^2_1)E(Y^2_3) - (E(Y_1))^2(E(Y_3))^2. \nonumber
\end{equation}
\begin{equation}
    Cov(Y_1,Y_1Y_3) = Var(Y_1)E(Y_3) = \frac{\kappa^1_2}{2}\kappa^{**}_1(e^{2\lambda t} - 1)(e^{\lambda t} - 1). \nonumber
\end{equation}
\begin{equation}
    Cov(Y_1,Y_1Y_2) = Var(Y_1)E(Y_2) = \frac{\kappa^1_2}{2}\kappa^{*}_1(e^{2\lambda t} - 1)(e^{\lambda t} - 1). \nonumber
\end{equation}
\begin{equation}
    Cov(Y^2_1,Y_1) = E(Y^3_1) - E(Y^2_1)E(Y_1).  \nonumber
\end{equation}
\begin{equation}
    Cov(Y_3,Y_1Y_3) = Var(Y_3)E(Y_1) = \frac{\kappa^{**}_2}{2}\kappa^{1}_1(e^{2\lambda t} - 1)(e^{\lambda t} - 1).   \nonumber
\end{equation}
\begin{equation}
    Cov(Y_2,Y_1Y_2) = Var(Y_2)E(Y_1) = \frac{\kappa^{*}_2}{2}\kappa^{1}_1(e^{2\lambda t} - 1)(e^{\lambda t} - 1).   \nonumber
\end{equation}
\begin{equation}
    Cov(Y^2_1,Y_1Y_3) = E(Y_3)(E(Y^3_1) - E(Y^2_1)E(Y_1)).   \nonumber
\end{equation}
\begin{equation}
    Cov(Y^2_1,Y_1Y_2) = E(Y_2)(E(Y^3_1) - E(Y^2_1)E(Y_1)).   \nonumber
\end{equation}
\begin{equation}
    Cov(Y_1Y_3,Y_1Y_2) = E(Y^2_1Y_2Y_3) - (E(Y_1Y_2)E(Y_1)E(Y_3)) = E(Y_2)E(Y_3)Var(Y_1) = \frac{\kappa^1_2}{2}\kappa^{**}_1\kappa^{*}_1(e^{2\lambda t} - 1)(e^{\lambda t} - 1)^2.
    \nonumber
\end{equation}
Using all the above covariances and variance terms in the \eqref{eq24} we can calculate the value of $Var[(\sigma_2)^2_t(\sigma_3)^2_t]$.

which concludes the proof of Theorem 6.$\qed$ \newline

Using \eqref{eq29} and considering $[X^{2}_T, X^{3}_T]$ of Equation \eqref{eq7} we substitute from Equation \eqref{eq22} \& \eqref{eq24} to get,

\begin{eqnarray}
  E [Cov(S^{3},S^{2})] &=& \frac{\gamma_{32}}{T}\int_{0}^{T}\bigg( \sqrt{E[(\sigma_2)^2_t(\sigma_3)^2_t]} - \frac{Var[(\sigma_2)^2_t(\sigma_3)^2_t]}{8(E[(\sigma_2)^2_t(\sigma_3)^2_t])^{\frac{3}{2}}}\bigg) dt  + \rho_{3}\rho_{2}\lambda \kappa^{1}_2 = E [Cov(S^{2},S^{3})]. \nonumber
\end{eqnarray}
Using \eqref{eq29} and considering $[X^{1}_T, X^{2}_T]$ of Equation \eqref{eq7} we substitute from Equation \eqref{eq25} \& \eqref{eq26} to get,

\begin{eqnarray}
  E [Cov(S^{1},S^{2})] &=& \frac{\gamma_{12}}{T}\int_{0}^{T}\bigg( \sqrt{E[(\sigma_1)^2_t(\sigma_2)^2_t]} - \frac{Var[(\sigma_1)^2_t(\sigma_2)^2_t]}{8(E[(\sigma_1)^2_t(\sigma_2)^2_t])^{\frac{3}{2}}}\bigg) dt  + \rho_{1}\rho_{2}\lambda \kappa^{1}_2 = E [Cov(S^{2},S^{1})]. \nonumber
\end{eqnarray}
Using \eqref{eq29} and considering $[X^{2}_T, X^{3}_T]$ of Equation \eqref{eq7} we substitute from Equation \eqref{eq27} \& \eqref{eq28} to get,

\begin{eqnarray}
  E [Cov(S^{1},S^{3})] &=& \frac{\gamma_{13}}{T}\int_{0}^{T}\bigg( \sqrt{E[(\sigma_1)^2_t(\sigma_3)^2_t]} - \frac{Var[(\sigma_1)^2_t(\sigma_3)^2_t]}{8(E[(\sigma_1)^2_t(\sigma_3)^2_t])^{\frac{3}{2}}}\bigg) dt  + \rho_{1}\rho_{3}\lambda \kappa^{1}_2 = E [Cov(S^{3},S^{1})]. \nonumber 
\end{eqnarray}

Collecting all the approximation results discussed above, we state another Theorem which will be used in the subsequent section.\newline

\begin{theorem}
The following are various results related to expected covariances and variances for multiple assets:\newline

\begin{enumerate}
  \item The expected covariance between $(S^{1},S^{2})$ is given by 
 \begin{eqnarray}
  E [Cov(S^{1},S^{2})] &=& \frac{\gamma_{12}}{T} \sum_{k=0}^{\infty}\sum_{p=0}^{k}\sum_{u=0}^{p} \frac{(-1)^{p+1}(2K)!}{4^{k}(k!)^2(2k-1)}\binom{k}{p}\binom{p}{u}(\beta_{12})^{-4p+2}(r_{2})^u(1 - (r_{2})^2)^{\frac{p - u}{2}} \nonumber \\&& \int_{0}^{T} e^{-\lambda t} N^{12}_t(p, u) dt  + \rho_{1}\rho_{2}\lambda \kappa^{1}_2 = E [Cov(S^{2},S^{1})]. \nonumber
 \end{eqnarray}
 
 \begin{eqnarray}
\textit{Or, }  E [Cov(S^{1},S^{2})] &=& \frac{\gamma_{12}}{T}\int_{0}^{T}\bigg( \sqrt{E[(\sigma_1)^2_t(\sigma_2)^2_t]} - \frac{Var[(\sigma_1)^2_t(\sigma_2)^2_t]}{8(E[(\sigma_1)^2_t(\sigma_2)^2_t])^{\frac{3}{2}}}\bigg) dt  + \rho_{1}\rho_{2}\lambda \kappa^{1}_2 = E [Cov(S^{2},S^{1})]. \nonumber \\&&
\end{eqnarray} 
  \item The expected covariance between $(S^{2},S^{3})$ is given by 
  \begin{eqnarray}
    E [Cov(S^{2},S^{3})] &=& \frac{\gamma_{23}}{T} \sum_{k=0}^{\infty}\frac{(-1)^{p+1}(2K)!}{4^{k}(k!)^2(2k-1)} \sum_{p=0}^{k}\binom{k}{p}\sum_{u=0}^{p}\binom{p}{u}\sum_{v=0}^{u}\binom{u}{v} \sum_{w=0}^{u-v}\binom{u}{v}\Big(\frac{r_3}{r_2}\Big)^{u-w}\Big(1 - r^2_2\Big)^{\frac{p-v-w}{2}} \nonumber \\&& \Big(1 - r^2_3\Big)^{\frac{p - u + w}{2}} (\beta_{23})^{- 4p +2}  \int_{0}^{T} e^{-\lambda t} N^{23}_t(p, u, v, w) dt  + \rho_{2}\rho_{3}\lambda \kappa^{1}_2 = E [Cov(S^{3},S^{2})]. \nonumber
\end{eqnarray}

\begin{eqnarray}
\textit{Or, }  E [Cov(S^{3},S^{2})] &=& \frac{\gamma_{32}}{T}\int_{0}^{T}\bigg( \sqrt{E[(\sigma_2)^2_t(\sigma_3)^2_t]} - \frac{Var[(\sigma_2)^2_t(\sigma_3)^2_t]}{8(E[(\sigma_2)^2_t(\sigma_3)^2_t])^{\frac{3}{2}}}\bigg) dt  + \rho_{3}\rho_{2}\lambda \kappa^{1}_2 = E [Cov(S^{2},S^{3})]. \nonumber \\&&
\end{eqnarray}
\item The expected covariance between $(S^{3},S^{1})$ is given by 
\begin{eqnarray}
    E [Cov(S^{3},S^{1})] &=& \frac{\gamma_{31}}{T} \sum_{k=0}^{\infty}\sum_{p=0}^{k}\sum_{u=0}^{p} \frac{(-1)^{p+1}(2K)!}{4^{k}(k!)^2(2k-1)}\binom{k}{p}\binom{p}{u}(\beta_{31})^{-4p+2}(r_{3})^u(1 - (r_{3})^2)^{\frac{p - u}{2}} \nonumber \\&& \int_{0}^{T} e^{-\lambda t} N^{31}_t(p, u) dt  + \rho_{3}\rho_{1}\lambda \kappa^{1}_2 = E [Cov(S^{1},S^{3})]. \nonumber
\end{eqnarray}

\begin{eqnarray}
\textit{Or, }  E [Cov(S^{1},S^{3})] &=& \frac{\gamma_{13}}{T}\int_{0}^{T}\bigg( \sqrt{E[(\sigma_1)^2_t(\sigma_3)^2_t]} - \frac{Var[(\sigma_1)^2_t(\sigma_3)^2_t]}{8(E[(\sigma_1)^2_t(\sigma_3)^2_t])^{\frac{3}{2}}}\bigg) dt  + \rho_{1}\rho_{3}\lambda \kappa^{1}_2 = E [Cov(S^{3},S^{1})]. \nonumber \\&&
\end{eqnarray}
\item The expected variance of $S^{1}$ is given by \begin{equation}
E [Var (S^{1})] = \frac{1}{T}\int_{0}^{T}e^{-\lambda t} N^{1}_t dt + \frac{(\rho_{1})^2}{T}\lambda \kappa^{1}_2  
\end{equation}
\item The expected variance of $S^{2}$ is given by \begin{equation}
 E [Var (S^{2})] = \frac{1}{T}\int_{0}^{T}e^{-\lambda t} N^{2}_t dt + \frac{(\rho_{2})^2}{T}\lambda \kappa^{1}_2. 
\end{equation}
\item The expected variance of $S^{3}$ is given by
\begin{equation}
E [Var (S^{3})] =  \frac{1}{T}\int_{0}^{T}e^{-\lambda t} N^{3}_t dt + \frac{(\rho_{3})^2}{T}\lambda \kappa^{1}_2. 
\end{equation}
\end{enumerate}
\end{theorem}

These results help us to derive the probability distribution of the eigenvalue that we discuss subsequently. We first look at the derivation of the trace swap price.

\section{Swap using the Trace and the largest Eigenvalue of the Covariance \\ Matrix}
\subsection{Swap using Trace}
As the first proposal, we consider the investor using the trace of the covariance matrix to develop the swap. The trace is given by 
\begin{equation}
    tr \hspace{2pt}\Omega =  Var(S^1) +  Var(S^2) +  Var(S^3)  \nonumber
\end{equation}

where 
\[ \Omega = 
\begin{bmatrix}
    Cov_R(S^1, S^1) & Cov_R(S^1, S^2) & Cov_R(S^1, S^3)  \\
    Cov_R(S^2, S^1) & Cov_R(S^2, S^2) & Cov_R(S^2, S^3) \\ Cov_R(S^3, S^1) & Cov_R(S^3, S^2) & Cov_R(S^3, S^3)  
\end{bmatrix}
\]
Now the price of the swap on trace is the expected present value of the payoff in the risk neutral world for the  assets we have considered

\begin{equation} 
    P_{trace} (x) = E \{e^{-rT}(tr \hspace{2pt}\Omega - K_{\text{strike price}})\}. \nonumber
\end{equation}
\begin{equation} 
    P_{trace}(x) = e^{-rT} E \{(tr \hspace{2pt}\Omega - K_{\text{strike price}})\}. \nonumber
\end{equation}
\begin{equation} 
    P_{trace}(x) = e^{-rT} E [tr \hspace{2pt}\Omega ] - e^{-rT} K_{\text{strike price}}. \nonumber
\end{equation}
\begin{equation} 
    P_{trace}(x) = e^{-rT} E [Var(S^1) +  Var(S^2) +  Var(S^3)  ] - e^{-rT} K_{\text{strike price}}. \nonumber
\end{equation}

Using the values from results 4, 5 and 6 from Theorem 7 we can write
\begin{eqnarray}\label{eq36}
    P_{trace}(x) &=& e^{-rT} \bigg[ \frac{1}{T}\int_{0}^{T}e^{-\lambda t} (\sum_{i = 1}^3 N^{i}_t) dt + \frac{\lambda \kappa^{1}_2}{T} (\sum_{i=1}^3 \rho_{i}^2) - K_{\text{strike price}} \bigg].
\end{eqnarray}

\subsection{Swap using the largest Eigenvalue}

The objective here is to define and derive the price of an eigenvalue swap. But we do not address the problem without an efficiency consideration as combinations of underlying assets for unconstrained variance may not be interesting as an investment destination.
So, here we assume that the investor considers the maximum eigenvalue of the covariance matrix, for a given expected mean return. For which we have to find the distribution. We are going to use the concept of quadratic optimization taken from the paper \citeauthor{Ref3}(\citeyear{Ref3}). Using the same notation, we execute the following constrained optimisation programme:

\begin{maxi}|l|
	  {\textbf{w(t)}}{\textbf{w(t)}^{T}\Omega \textbf{w(t)}}{}{}
	  \addConstraint{\textbf{w(t)}^{T}\textbf{w(t)}}{=1}{}
	  \addConstraint{\textbf{I}^T \textbf{w(t)}}{= 1}{}
	  \addConstraint{\text{E}(\textbf{R})^T\textbf{w(t)}}{= k,}{}\nonumber
\end{maxi}
where \textbf{w(t)} is the weight vector corresponding to each of the stocks and \textbf{R} is the vector containing the expected return of the stocks. Overall the constraint can be combined as 
\begin{maxi}|l|
	  {\textbf{w(t)}}{\textbf{w(t)}^{T}\Omega \textbf{w(t)}}{}{}
	  \addConstraint{\textbf{w(t)}^{T}\textbf{I} \textbf{w(t)}}{=1}{}
	  \addConstraint{\textbf{A}^T \textbf{w(t)}}{= \bf{b},}{}\nonumber
\end{maxi}
where 
\[ \textbf{w(t)} = 
\begin{bmatrix}
    w_1(t) \\ w_2(t) \\w_3(t) 
 \end{bmatrix}
\]
\[ \textbf{A} = 
\begin{bmatrix}
    \text{E}(\textbf{R}) & \textbf{I} 
 \end{bmatrix}
\]
\[ \bf{b} = 
\begin{bmatrix}
   k \\
    1 
\end{bmatrix}.
\]

In our situation we have three assets so,
\[ \Omega = 
\begin{bmatrix}
    Cov_R(S^1, S^1) & Cov_R(S^1, S^2) & Cov_R(S^1, S^3)  \\
    Cov_R(S^2, S^1) & Cov_R(S^2, S^2) & Cov_R(S^2, S^3) \\ Cov_R(S^3, S^1) & Cov_R(S^3, S^2) & Cov_R(S^3, S^3)  
\end{bmatrix}
\]

\[ \text{and }
\textbf{A}=
\begin{bmatrix}
    0 & 1 \\
    \mu_2 & 1\\
    \mu_3 & 1 \\
\end{bmatrix}.
\]

We have considered mean of one asset as 0, now doing a QR decomposition of the matrix \textbf{A} as defined in the paper by \citeauthor{Ref4} (\citeyear{Ref4}), we get
\[
\textbf{P} = \begin{bmatrix}
\textbf{P}_1 & \textbf{P}_2
\end{bmatrix},
\]
\[ \text{where }
\textbf{P}_1=
\begin{bmatrix}
    0 & \frac{\mu^2_2 + \mu^2_3}{\sqrt{2(\mu^2_2 + \mu^2_3 - \mu_2\mu_3)(\mu^2_2 + \mu^2_3)}} \\
   \frac{\mu_2}{\sqrt{\mu^2_2 + \mu^2_3}} & \frac{ \mu^2_3 - \mu_2\mu_3}{\sqrt{2(\mu^2_2 + \mu^2_3 - \mu_2\mu_3)(\mu^2_2 + \mu^2_3)}}\\
   \frac{\mu_3}{\sqrt{\mu^2_2 + \mu^2_3}} & \frac{\mu^2_2 - \mu_3\mu_2}{\sqrt{2(\mu^2_2 + \mu^2_3 - \mu_2\mu_3)(\mu^2_2 + \mu^2_3)}}
   
\end{bmatrix}
\]

\[ \text{and }
\textbf{P}_2=
\begin{bmatrix}
     \frac{\mu_2 - \mu_3}{\sqrt{2(\mu^2_2 + \mu^2_3 - \mu_2\mu_3)}} \\
   \frac{\mu_3}{\sqrt{2(\mu^2_2 + \mu^2_3 - \mu_2\mu_3)}}\\
   -\frac{\mu_2}{\sqrt{2(\mu^2_2 + \mu^2_3 - \mu_2\mu_3)}}
   
\end{bmatrix}.
\]

\[\textbf{A} = 
\begin{bmatrix}
\textbf{P}_1 & \textbf{P}_2
\end{bmatrix}
\begin{bmatrix}
\textbf{R} \\ 0
\end{bmatrix}
\]

\[ \text{where }
\textbf{R}=
\begin{bmatrix}
    \sqrt{\mu^2_2 + \mu^2_3} & \frac{\mu_2 + \mu_3}{\sqrt{\mu^2_2 + \mu^2_3}} \\
   0 & \sqrt{\frac{2(\mu^2_2 + \mu^2_3 - \mu_2\mu_3 )}{\mu^2_2 + \mu^2_3}}\\
\end{bmatrix}.
\]

Let us consider the following definitions,
\begin{equation}
%y-U
\sqrt{\mu^2_2 + \mu^2_3} = U. \nonumber
\end{equation}
\begin{equation}
\sqrt{2(\mu^2_2 + \mu^2_3 - \mu_2\mu_3)(\mu^2_2 + \mu^2_3)} = W. \nonumber
\end{equation}
\begin{equation}
 \sqrt{2(\mu^2_2 + \mu^2_3 - \mu_2\mu_3)} = Z. \nonumber   
\end{equation}
\begin{equation}
  \mu_2 - \mu_3 = V. \nonumber  
\end{equation}

We finally want to compute
\[
\begin{bmatrix}
\textbf{P}_1 & \textbf{P}_2
\end{bmatrix}^{T} 
\begin{bmatrix}
    Cov_R(S^1, S^1) & Cov_R(S^1, S^2) & Cov_R(S^1, S^3)  \\
    Cov_R(S^2, S^1) & Cov_R(S^2, S^2) & Cov_R(S^2, S^3) \\ Cov_R(S^3, S^1) & Cov_R(S^3, S^2) & Cov_R(S^3, S^3)  
\end{bmatrix}
\begin{bmatrix}
\textbf{P}_1 & \textbf{P}_2
\end{bmatrix}.
\]

\begin{theorem}
From the paper (\citeauthor{Ref3}(\citeyear{Ref3})) we use value of the matrix \textbf{F} 
\[\textbf{F} = 
\begin{bmatrix}
      \frac{1}{\det R}\big(\frac{k Z}{U}\big) \\
   \frac{1}{\det R}\big(-k\big(\frac{\mu_2 + \mu_3}{Y}\big) + U\big) \\ \sqrt{1 - \frac{1}{(\det R)^2} \big(k^2 \frac{Z^2}{U^2} + 
  \big(\big(-k\frac{\mu_2 + \mu_3}{U}\big) + U\big)^2\big)}
\end{bmatrix}
\]
which is used to calculate $\textbf{w(t) = PF}$
\end{theorem}
Now using this theorem 8, we can write the maximum eigenvalue as,

\begin{equation}
    \lambda = \textbf{w(t)}^{T}\Omega \textbf{w(t)}. \nonumber
\end{equation}
\begin{equation} \label{eq37}
    \lambda = \textbf{F}^{T}\textbf{P}^{T}\Omega\textbf{P}\textbf{F}.
\end{equation}

With the previous definition we can simplify \textbf{P} as
\[
\textbf{P}=
\begin{bmatrix}
    0 & \frac{\mu^2_2 + \mu^2_3}{W} & \frac{\mu_2 - \mu_3}{Z} \\
   \frac{\mu_2}{U} & \frac{\mu^2_3 - \mu_2\mu_3}{W} &  \frac{\mu_3}{Z}\\
   \frac{\mu_3}{U} & \frac{\mu^2_2 - \mu_3\mu_2}{W} & -\frac{\mu_2}{Z}
\end{bmatrix} = 
\begin{bmatrix}
    0 & \frac{U^2}{W} & \frac{V}{Z} \\
   \frac{\mu_2}{U} & -\frac{V\mu_3}{W} &  \frac{\mu_3}{Z}\\
   \frac{\mu_3}{U} & \frac{V\mu_2}{W} & -\frac{\mu_2}{Z}
\end{bmatrix}
\]
Then, $\textbf{P}^{T}\Omega\textbf{P} = $
\[
\begin{bmatrix}
     0 & \frac{U^2}{W} & \frac{V}{Z} \\
   \frac{\mu_2}{U} & -\frac{V\mu_3}{W} &  \frac{\mu_3}{Z}\\
   \frac{\mu_3}{U} & \frac{V\mu_2}{W} & -\frac{\mu_2}{Z}
\end{bmatrix}^{T}
\begin{bmatrix}
    Cov_R(S^1, S^1) & Cov_R(S^1, S^2) & Cov_R(S^1, S^3)  \\
    Cov_R(S^2, S^1) & Cov_R(S^2, S^2) & Cov_R(S^2, S^3) \\ Cov_R(S^3, S^1) & Cov_R(S^3, S^2) & Cov_R(S^3, S^3)  
\end{bmatrix}
\begin{bmatrix}
    0 & \frac{U^2}{W} & \frac{V}{Z} \\
   \frac{\mu_2}{U} & -\frac{V\mu_3}{W} &  \frac{\mu_3}{Z}\\
   \frac{\mu_3}{U} & \frac{V\mu_2}{W} & -\frac{\mu_2}{Z}
\end{bmatrix} = \begin{bmatrix}
       \textbf{g}_1 & \textbf{g}_2 &\textbf{g}_3
\end{bmatrix}
\]
Multiplying out, we get the individual vectors $\textbf{g}_1$ etc. as
\[\textbf{g}_1 =
\begin{bmatrix}
    \frac{\mu_2}{U}\big(\frac{\mu_2}{U}Cov(S^2, S^2) + \frac{\mu_3}{U}Cov(S^3, S^2)\big) + \frac{\mu_3}{U}\big(\frac{\mu_2}{Y}Cov(S^2, S^3) + Cov(S^3, S^3)\big) \\[2.5ex] \frac{\mu_2}{U}\big(\frac{U^2}{W}Cov(S^1, S^2) - \frac{V\mu_3}{W}Cov(S^2, S^2) + \frac{V\mu_2}{W}Cov(S^3, S^2)\big) + \frac{\mu_3}{U}\big(\frac{U^2}{X}Cov(S^1, S^3) - \frac{V\mu_3}{W}Cov(S^2, S^3) + \frac{V\mu_2}{W}Cov(S^3, S^3)\big) \\[2.5ex]  \frac{\mu_2}{U}\big(\frac{V}{Z}Cov(S^1, S^2) + \frac{\mu_3}{Z}Cov(S^2, S^2) - \frac{\mu_2}{Z}Cov(S^3, S^2)\big) + \frac{\mu_3}{U}\big(\frac{V}{Z}Cov(S^1, S^3) + \frac{\mu_3}{Z}Cov(S^2, S^3) - \frac{\mu_2}{Z}Cov(S^3, S^3)\big) 
\end{bmatrix}
\]
\[ \textbf{g}_2 = 
 \begin{bmatrix}
 \begin{matrix}
  \frac{U^2}{W}\big(\frac{\mu_2}{U}Cov(S^2, S^1) + \frac{\mu_3}{U}Cov(S^3, S^1)\big) -  \frac{V\mu_3}{W}\big(\frac{\mu_2}{U}Cov(S^2, S^2) + \frac{\mu_3}{U}Cov(S^3, S^2)\big) + \frac{V\mu_2}{W}\big(\frac{\mu_2}{U}Cov(S^2, S^3) + \frac{\mu_3}{U}Cov(S^3, S^3)\big) 
  \end{matrix}
  \\[5ex] 
  \begin{matrix}
  \frac{U^2}{W}\big(\frac{U^2}{W}Cov(S^1, S^1)- \frac{V\mu_3}{W}Cov(S^2, S^1) + \frac{V\mu_2}{W}Cov(S^3, S^1)\big) - \\ \hfill{} \frac{V\mu_3}{W}\big(\frac{U^2}{W}Cov(S^1, S^2) - \frac{V\mu_3}{W}Cov(S^2, S^2) + \frac{V\mu_2}{W}Cov(S^3, S^2)\big) + \frac{V\mu_2}{W}\big(\frac{U^2}{W}Cov(S^1, S^3) - \frac{V\mu_3}{W}Cov(S^2, S^3) + \frac{V\mu_2}{W}Cov(S^2, S^1)\big)
  \end{matrix}
  \\[5ex]
  \begin{matrix}
    \frac{V}{Z}\big(\frac{V}{Z}Cov(S^1, S^1) - \frac{V\mu_3}{W}Cov(S^2, S^1) + \frac{V\mu_2}{W}Cov(S^3, S^1)\big) + \\ \hfill{} \frac{\mu_3}{Z}\big(\frac{V}{Z}Cov(S^1, S^2) + \frac{\mu_3}{Z}Cov(S^2, S^2) - \frac{\mu_2}{Z}Cov(S^3, S^2)\big) - \frac{\mu_2}{Z}\big(\frac{V}{Z}Cov(S^1, S^3) + \frac{\mu_3}{Z}Cov(S^2, S^3) - \frac{\mu_2}{Z}Cov(S^3, S^3)\big) 
    \end{matrix}
    \end{bmatrix}
    \]
 \[ \textbf{g}_3 = 
    \begin{bmatrix}
\begin{matrix}
\frac{V}{Z}\big(\frac{\mu_2}{U}Cov(S^2, S^1) + \frac{\mu_3}{U}Cov(S^3, S^1)\big) + \frac{\mu_3}{Z}\big(\frac{\mu_2}{U}Cov(S^2, S^2) + \frac{\mu_3}{U}Cov(S^3, S^2)\big)  - \frac{\mu_2}{W}\big(\frac{\mu_2}{U}Cov(S^2, S^3) + \frac{\mu_3}{U}Cov(S^3, S^3)\big)   \end{matrix}        \\[5ex]
\begin{matrix}
\frac{V}{Z}\big(\frac{U^2}{W}Cov(S^1, S^1) - \frac{V\mu_3}{W}Cov(S^2, S^1) + \frac{V\mu_2}{W}Cov(S^3, S^1)\big) +\\ \hfill{}  \frac{\mu_3}{Z}\big(\frac{U^2}{W}Cov(S^1, S^2) - \frac{V\mu_3}{W}Cov(S^2, S^2) + \frac{V\mu_2}{W}Cov(S^3, S^2)\big) - \frac{\mu_2}{Z}\big(\frac{U^2}{W}Cov(S^1, S^3) - \frac{V\mu_3}{W}Cov(S^2, S^3) + \frac{V\mu_2}{W}Cov(S^3, S^3)\big) 
\end{matrix}\\[5ex]
\begin{matrix}
        \frac{V}{Z}\big(\frac{V}{Z}Cov(S^1, S^1) + \frac{\mu_3}{Z}Cov(S^2, S^1) - \frac{\mu_2}{Z}Cov(S^3, S^1)\big) + \\ \hfill{} \frac{\mu_3}{Z}\big(\frac{V}{Z}Cov(S^1, S^2) + \frac{\mu_3}{Z}Cov(S^2, S^2) - \frac{\mu_2}{Z}Cov(S^3, S^2)\big) - \frac{\mu_2}{Z}\big(\frac{U^2}{W}Cov(S^1, S^3) + \frac{\mu_3}{Z}Cov(S^2, S^3) - \frac{\mu_2}{Z}Cov(S^3, S^3)\big)
 \end{matrix}  
\end{bmatrix}
\]
\\
So, using \eqref{eq37} we can calculate the eigenvalue as, 
\[ \lambda = 
\textbf{F}^{T}  
\begin{bmatrix}
 \textbf{g}_1 & \textbf{g}_2 & \textbf{g}_3   
\end{bmatrix}
\textbf{F}
\]

where let us assign
\[
\textbf{F}^{T}  
\begin{bmatrix}
 \textbf{g}_1 & \textbf{g}_2 & \textbf{g}_3   
\end{bmatrix} = 
\begin{bmatrix}
 \textbf{d}_1 & \textbf{d}_2 & \textbf{d}_3   
\end{bmatrix}
\]
\begin{eqnarray}
  \textbf{d}_1 &=& \bigg(\frac{\mu^2_2}{U^2}Cov(S^2,S^2) + 2\frac{\mu_3\mu_2}{U^2}Cov(S^3,S^2) +  \frac{\mu^2_3}{U^2}Cov(S^3,S^3)\bigg)\bigg(\frac{1}{\det R}\Big(\frac{k Z}{U}\Big)\bigg) + \bigg(\frac{U\mu_2}{W}Cov(S^1,S^2) +  \nonumber \\&& \frac{V}{WU}Cov(S^3,S^2)\Big(\mu^2_2 - \mu^2_3\Big) + \frac{U\mu_3}{W}Cov(S^1,S^3)  \frac{V\mu_2\mu_3}{WU}\Big(Cov(S^3,S^3) - Cov(S^2,S^2)\Big)\bigg)\bigg(\frac{1}{\det R}\Big(-k\Big(\frac{\mu_2 + \mu_3}{U}\Big) + U\Big)\bigg) + \nonumber \\&&   \bigg(\frac{V\mu_2}{UZ}Cov(S^1,S^2) +  \frac{\mu_3\mu_2}{UZ}\Big(Cov(S^2,S^2) - Cov(S^3,S^3)\Big) + \frac{V\mu_3}{ZU}Cov(S^1,S^3) + \frac{1}{UZ}Cov(S^2,S^3)\Big(\mu^2_3 - \mu^2_2\Big) \bigg)   \nonumber \\&& \Bigg(\sqrt{1 - \frac{1}{(\det R)^2} \bigg(k^2 \frac{Z^2}{U^2} + 
  \Big(\Big(-k\frac{\mu_2 + \mu_3}{U}\Big) + U\Big)^2\bigg)}\Bigg).  \nonumber
  \end{eqnarray}

\begin{eqnarray}
  \textbf{d}_2 &=& 
  \bigg(\frac{U\mu_2}{W}Cov(S^2,S^1) + \frac{U\mu_3}{W}Cov(S^3,S^1) + \frac{V\mu_2\mu_3}{WU}\Big(Cov(S^3,S^3) -Cov(S^2,S^2)\Big) + \frac{V}{WU}Cov(S^3,S^2) \Big(\mu^2_2 - \mu^2_3\Big)\bigg)\bigg(\frac{1}{\det R}\Big(\frac{k Z}{U}\Big)\bigg)  \nonumber \\&& +  \bigg(\frac{U^4}{W^2}Cov(S^1,S^1) - 2\frac{U^2V\mu_3}{W^2}Cov(S^2,S^1) + 2\frac{U^2V\mu_2}{W^2}Cov(S^3,S^1) + \frac{V^2\mu^2_3}{W^2}Cov(S^2,S^2) - 2\frac{V^2\mu_2\mu_3}{W^2}Cov(S^3,S^2) + \nonumber \\&& \frac{V^2\mu^2_2}{W^2}Cov(S^3,S^3)\bigg) \bigg(\frac{1}{\det R}\big(-k\Big(\frac{\mu_2 + \mu_3}{U}\Big) + U\Big)\bigg) + 
   \Big(\frac{V^2}{Z^2}Cov(S^1,S^1) + \frac{V^2\mu_2}{WZ}\Big(\frac{V}{W} - \frac{1}{Z}\Big)Cov(S^3,S^1) +  \frac{V\mu_3}{Z}\Big(\frac{1}{Z} - \frac{V}{W}\Big)\nonumber \\&&Cov(S^1,S^2) +   \frac{\mu^2_3}{Z^2}Cov(S^2,S^2) - 2\frac{\mu_3\mu_2}{Z^2}Cov(S^2,S^3) + \frac{\mu^2_2}{Z^2}Cov(S^3,S^3)\bigg)\Bigg(\sqrt{1 - \frac{1}{(\det R)^2} \bigg(k^2 \frac{Z^2}{U^2} + 
  \Big(\Big(-k\frac{\mu_2 + \mu_3}{U}\Big) + U\Big)^2\bigg)}\Bigg). \nonumber
\end{eqnarray}

\begin{eqnarray}
   \textbf{d}_3 &=&
   \bigg(\frac{V\mu_2}{ZU}Cov(S^2,S^1) + \frac{V\mu_3}{ZU}Cov(S^3,S^1) +  \frac{\mu_2\mu_3}{ZU}Cov(S^2,S^2) + \frac{\mu^2_3}{ZU}Cov(S^3,S^2)  - \frac{\mu^2_2}{UW}Cov(S^2,S^3) - \frac{\mu_3\mu_2}{WU}Cov(S^3,S^3)\bigg) \nonumber \\&&  \bigg(\frac{1}{\det R}\Big(\frac{k Z}{U}\Big)\bigg) + \bigg(\frac{VU^2}{ZW}Cov(S^1,S^1) + \frac{\mu_3}{ZW}Cov(S^2,S^1) (U^2 - V^2) + \frac{\mu_2}{WZ}Cov(S^3,S^1)(V^2 - U^2)  - \frac{V\mu^2_3}{WZ}Cov(S^2,S^2) + \nonumber \\&& 2\frac{V\mu_2\mu_3}{WZ}Cov(S^3,S^2) -  \frac{V\mu^2_2}{WZ}Cov(S^3,S^3)\bigg)\bigg(\frac{1}{\det R}\big(-k\Big(\frac{\mu_2 + \mu_3}{U}\Big) + U\Big)\bigg) + 
    \bigg(\frac{V^2}{Z^2}Cov(S^1,S^1) + 2\frac{V\mu_3}{Z^2}Cov(S^2,S^1) - \nonumber \\&& Cov(S^3,S^1) +  \frac{\mu^2_3}{Z^2}Cov(S^2,S^2) - 2\frac{\mu_2\mu_3}{Z^2}Cov(S^3,S^2) -  \frac{\mu_2}{Z}Cov(S^1,S^3)\Big(\frac{U^2}{W} + \frac{V}{Z}\Big) + \frac{\mu^2_2}{Z^2}Cov(S^3,S^3)\bigg) \nonumber \\&& \Bigg(\sqrt{1 - \frac{1}{(\det R)^2} \Big(k^2 \frac{Z^2}{U^2} +
  \big(\big(-k\frac{\mu_2 + \mu_3}{U}\big) +  U\big)^2\Big)}\Bigg). \nonumber
\end{eqnarray}

\begin{eqnarray}\label{eq38}
   \lambda &=& \textbf{d}_1 \Big(\frac{1}{\det R}\big(\frac{k Z}{U}\big)\Big) + \textbf{d}_2 \Big(\frac{1}{\det R}\big(-k\big(\frac{\mu_2 + \mu_3}{U}\big) + Y\big)\Big) + \textbf{d}_3 \sqrt{1 - \frac{1}{(\det R)^2} \Big(k^2 \frac{Z^2}{U^2} + \big(\big(-k\frac{\mu_2 + \mu_3}{U}\big) +  U\big)^2\Big)}. \nonumber \\&&
\end{eqnarray}
Now the price of the largest eigenvalue swap is the expected present value of the payoff in the risk neutral world for this three asset we have considered

\begin{equation} 
    P_{eigenvalue} = E \{e^{-rT}(\lambda  - K)\}. \nonumber
\end{equation}
\begin{equation} 
    P_{eigenvalue} = e^{-rT} E \{(\lambda  - K)\}. \nonumber
\end{equation}
Replacing the Equation \eqref{eq38} in to the above equation,
\begin{eqnarray} 
    P_{eigenvalue} &=& e^{-rT} E \bigg(\textbf{d}_1 \Big(\frac{1}{\det R}\big(\frac{k Z}{U}\big)\Big) + \textbf{d}_2 \Big(\frac{1}{\det R}\big(-k\big(\frac{\mu_2 + \mu_3}{U}\big) + Y\big)\Big) + \nonumber \\&& \textbf{d}_3 \sqrt{1 - \frac{1}{(\det R)^2} \Big(k^2 \frac{Z^2}{U^2} + \big(\big(-k\frac{\mu_2 + \mu_3}{U}\big) +  U\big)^2\Big)}\bigg) -  \big(e^{-rT} \times K_{strike price}\big). \nonumber
\end{eqnarray}
\begin{eqnarray} \label{eq39}
    P_{eigenvalue} &=& e^{-rT}  \bigg(E(\textbf{d}_1) \Big(\frac{1}{\det R}\big(\frac{k Z}{U}\big)\Big) + E(\textbf{d}_2) \Big(\frac{1}{\det R}\big(-k\big(\frac{\mu_2 + \mu_3}{U}\big) + Y\big)\Big) + \nonumber \\&& E(\textbf{d}_3) \sqrt{1 - \frac{1}{(\det R)^2} \Big(k^2 \frac{Z^2}{U^2} + \big(\big(-k\frac{\mu_2 + \mu_3}{U}\big) +  U\big)^2\Big)} - K_{strike price}\bigg). 
\end{eqnarray}

Using the values (1-6) obtained in Theorem 7, we can calculate the values of $E(\textbf{d}_1), E(\textbf{d}_2)$ and $E(\textbf{d}_3)$ which will finally give the $P_{eigenvalue}$ in \eqref{eq39}.

\section{Numerical Example}

Spot Price of three agricultural commodities are considered for our numerical illustration\footnote{We thank Suranjana Joarder for help with data access.}. The first one is Potato  (\citeauthor{Quantopian}, \citeyear{Quantopian}), spot Price of Mustard is chosen as the second commodity and spot price of Rice is chosen as third. The data is taken from the Multi-Commodity Exchange of India and we have used the daily closing price (which we have considered the Spot price) of these three commodities, denoted by $S_{1}$, $S_{2}$ and $S_{3}$ respectively, in the time range 1st June, 2019 till 31st May, 2020. We computed the return of the spot prices using the formula
\begin{equation}
    \text{return at time t:} r_{it} = \frac{S_{it} - S_{i,t-1}}{S_{i,t-1}}, \; i = 1, 2, 3. \nonumber
\end{equation}

We estimate the relevant parameters from the descriptive statistics for the three commodities presented in the table below. From the statistics we can observe that range of the returns of Mustard is least whereas that of Rice is highest. Also the median value of the returns of the three commodities are not close to the mean showing that the data is skewed in nature. The volatility  of  Rice is least whereas for Potato it is maximum which goes with the intuition that rice has less volatility in the market. \newline
\begin{center}
{\bf Table 1: Descriptive Statistics}
\begin{tabular}{| m{3cm} | m{3cm} | m{3cm} | m{3cm} |} 
\hline
Commodity & Mustard & Potato & Rice \\
\hline
No. of Observation & 252 & 252 & 252\\
\hline
Mean & -0.0038  & 0.0317 & -0.0002\\
\hline
Std. Error & 0.052 & 0.044 & 0.027\\ 
\hline
95\% CI of Mean  & 0.095 & 0.0205 & 0.0115\\
\hline
Min & -0.0526 &  -0.0916 & -0.990\\
\hline
Max & 0.01465 & 0.0618 & 0.0432 \\
\hline
Range & 0.0941 & 0.1598 & 1.033\\
\hline
Median & 0 & 0.0005 & 0.0018\\
\hline
Std. Deviation & 0.0502 & 0.0267 & 0.0058\\
\hline
\end{tabular}
\end{center}

\begin{figure}[H]
\centering
\begin{subfigure}{.6\textwidth}
  \centering
  \includegraphics[width=0.85\textwidth]{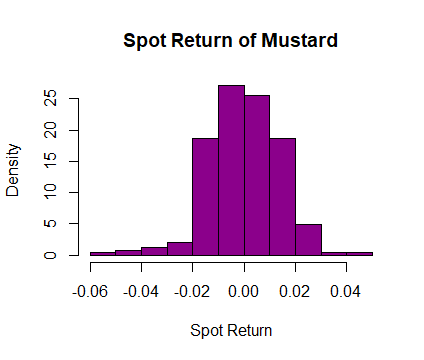}
   \caption{Spot Return of Mustard}
  \label{fig:1}
\end{subfigure}%
\begin{subfigure}{.6\textwidth}
  \centering
  \includegraphics[width=0.85\textwidth]{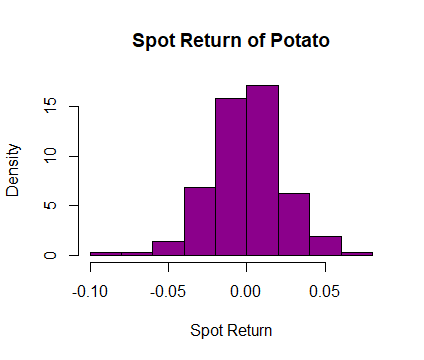}
\caption{Spot Return of Potato }
  \label{fig:2}
\end{subfigure}
\begin{subfigure}{.6\textwidth}
  \centering
  \includegraphics[width=0.85\textwidth]{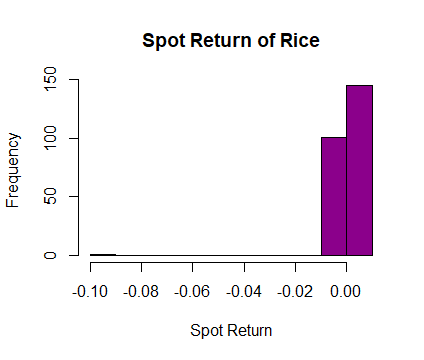}
   \caption{Spot Return of Rice }
  \label{fig:3}
\end{subfigure}%
\caption{Histogram of the Spot Return for the 3 commodities}
\label{fig:test}
\end{figure}

Plotting the histogram of the return for Mustard, Potato and Rice we can see that for Mustard and Potato it is evenly spread [Refer Figure (\ref{fig:1} \& \ref{fig:2})] but for Rice it is highly skewed [Refer Figure (\ref{fig:3})].

\begin{figure}[H]
\centering
\includegraphics[width=0.95\textwidth]{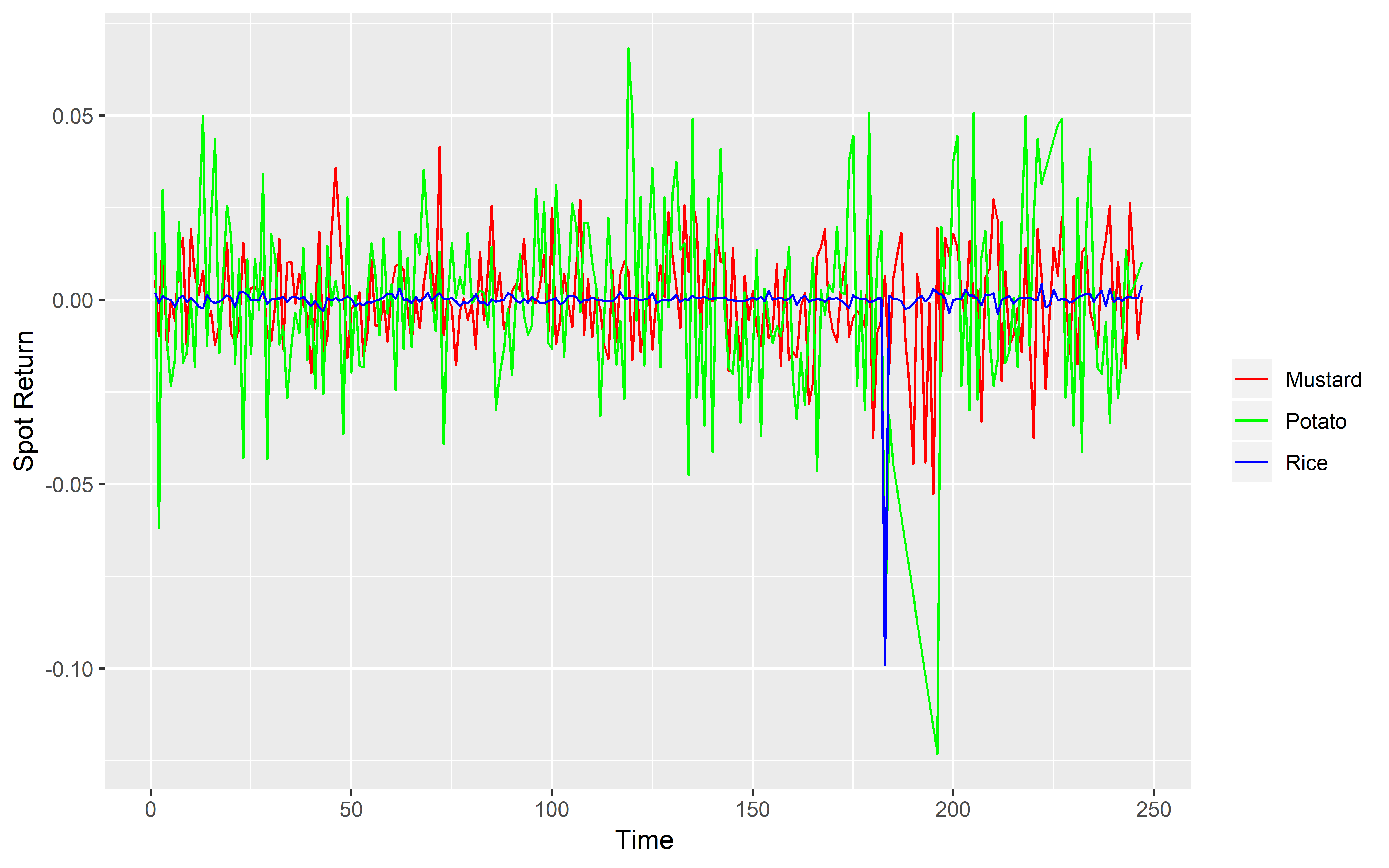}
\caption{Line Plots of the Spot Return for the 3 commodities}
\label{fig:8}
\end{figure}

Plotting the line diagrams of the spot return for the three commodities over the entire time period we can see there are sudden movements indicating that there are clear Jumps in all the commodities [Refer Figure \ref{fig:8}].\newline
Below we also show the required parameter estimates, derived from the descriptive statistics, which are going to be used in the pricing calculations. 

\begin{itemize}
    \item 
    The mean, standard deviation and leverage parameters of returns of the commodities are estimated as (standard error of the estimate in bracket),

\begin{tabular}{| m{1cm} | m{3cm} | } 
\hline
$\mu_1$ & -0.0038 (0.052)\\
\hline
$\mu_2$ & 0.0317 (0.044)\\ 
\hline
$\mu_3$ & -0.0002 (0.027)\\ 
\hline
\end{tabular}
\quad
\begin{tabular}{| m{1cm} | m{3cm} | } 
\hline
$(\sigma_1)_0$ & 0.0502 (0.0643)\\
\hline
$(\sigma_2)_0$  & 0.0267 (0.0761)\\ 
\hline
$(\sigma_3)_0$ & 0.0058 (0.0097)\\ 
\hline
\end{tabular}
\quad
\begin{tabular}{| m{1cm} | m{2cm} | } 
\hline
$\rho_{1}$  & 0.8 (0.065)\\
\hline
$\rho_{2}$  & 0.5 (0.034)\\ 
\hline
$\rho_{3}$  & 0.6 (0.087)\\ 
\hline
\end{tabular}

\item 
   From the paper by \citeauthor{Lonnbark} (\citeyear{Lonnbark}) we can calculate the correlation of returns among the  commodities as,

\begin{center}
\begin{tabular}{| m{1cm} | m{2cm} | } 
\hline
$\gamma_{12}$ & -0.0216\\
\hline
$\gamma_{23}$ & -0.0862\\ 
\hline
$\gamma_{31}$ & -0.0276\\ 
\hline
\end{tabular}
\end{center}

\item 
    The correlation value of the L\'evy subordinators and $\lambda$ of the  commodities are calculated as (\citeauthor{Jacod}(\citeyear{Jacod})),

\begin{center}
\begin{tabular}{| m{1cm} | m{2cm} | } 
\hline
$r_{2}$ & 0.2319\\
\hline
$r_{3}$ & 0.5721\\ 
\hline
$\lambda$ & 0.4\\ 
\hline
$r$ & 0.00014\\
\hline
\end{tabular}
\end{center}
\end{itemize}

We can evaluate the expected value of the variance term and the covariance term using Theorem 7 and the above values to obtain the matrix as:

\[
\begin{bmatrix}
    E[(Var(S^1, S^1)] & E[Cov_R(S^1, S^2)] & E[Cov_R(S^1, S^3)]  \\
    E[Cov_R(S^2, S^1)] & E[Var(S^2, S^2)] & E[Cov_R(S^2, S^3)] \\ E[Cov_R(S^3, S^1)] & E[Cov_R(S^3, S^2)] & E[Var(S^3, S^3)]  
\end{bmatrix}
=
\begin{bmatrix}
    0.00736 & 0.00065 & 0.00082  \\
    0.00065 & 0.00498 & 0.00039 \\ 0.00082 & 0.00039 & 0.00217 
\end{bmatrix}
\]

As mentioned in the introduction, our first candidate measure of generalized variance is the trace of the covariance matrix which is nothing but the sum of the individual variances. Intuitively, this is the variance of the return of a portfolio comprising one unit of each of the stocks, assuming them to be uncorrelated. So we can calculate the price of the swap given by Equation \eqref{eq36} as,

\begin{equation} 
    P_{trace}(x) = e^{-rT}\times 0.01451 - e^{-rT} K_{\text{strike price}} \nonumber
\end{equation}
The swap is written of the trace. Considering the strike price as 0.01 and duration is for 1 year i.e. T = 252 we can finally calculate the trace swap as,

\begin{equation} 
    P_{\text{trace}} = 0.00435 \nonumber
\end{equation}

The second candidate measure of generalised variance considered here is the maximum eigenvalue which is the magnitude of the biggest component of the orthogonalised system for the return covariance matrix. As the return distributions are correlated (the covariance matrix is not diagonal), this biggest component will be significantly larger than the individual variances. This is considered as we are interested in managing the variance, so swapping for the biggest component is a safe strategy to adopt.

Considering T = 252 we do the following calculations,
\[
\textbf{A}=
\begin{bmatrix}
    -0.0038 & 1 \\
    0.0317 & 1\\
    -0.0002 & 1 \\
\end{bmatrix}
\]
Doing a QR decomposition of A, we get
\[
\textbf{P}_1=
\begin{bmatrix}
    -0.119 & 0.7359 \\
    0.9929 & 0.0925\\
    -0.0063 & 0.6707
\end{bmatrix}
\]
\[
\textbf{R}=
\begin{bmatrix}
    0.0319 & 0.8676 \\
    0 & 1.4991\\
\end{bmatrix}
\]
\[
\textbf{P}_2=
\begin{bmatrix}
    0.6874  \\
    0.0747 \\
    0.7417 
\end{bmatrix}
\]
Therefore P is,
\[
\textbf{P}=
\begin{bmatrix}
    -0.119 & 0.7359 & 0.6874\\
    0.9929 & 0.0925 & 0.0747\\
    -0.0063 & 0.6707 & 0.7417
\end{bmatrix}
\]
\[
\textbf{R}^{-T}=
\begin{bmatrix}
    31.3479 & 0 \\
    -18.1425 & 0.6670\\
\end{bmatrix}
\]
Given that,

\[ \bf{b} = 
\begin{bmatrix}
   0.0007 \\
    1 
\end{bmatrix}
\]

\[
\textbf{q} = 
\begin{bmatrix}
     0.0219 \\
  0.6543
\end{bmatrix}
\]
\[
\textbf{r} = 
\begin{bmatrix}
     0.7559
\end{bmatrix}
\]
Therefore we can calculate \textbf{w(t)}
\[
\textbf{w(t)} = 
\begin{bmatrix}
       0.9985 \\
  0.1387 \\  0.9994
\end{bmatrix}
\]

Using the Equation \eqref{eq39} we can calculate the price of the eigenvalue swap is given by,
\begin{equation} 
    P_{\text{eigenvalue}} = e^{-rT} 0.0115
- \big(e^{-rT} \times K_{strike price}\big) \nonumber
\end{equation}

The swap is written of the eigenvalue. Considering the strike price as 0.01 we can finally calculate the eigenvalue swap as,

\begin{equation} 
    P_{\text{eigenvalue}} = 0.00145 \nonumber
\end{equation}

\section{Conclusion}

In this paper we have presented a new approach for pricing swaps for a Barndorff-Nielsen and Shephard model defined on two important measures of generalized variance,  namely the maximum eigenvalue and trace of the covariance matrix of the returns on assets involved. The objective is to price generalized variance swaps for financial markets guided by the BNS model for asset returns. We have considered multiple assets in the portfolio for  theoretical purpose and demonstrated the theoretical approach with the help of numerical examples taking three commodities in the portfolio. The results derived in this paper are the comparison between the swaps defined by the trace and the eigenvalue. Moreover, the results obtained in this paper have important implications for their use in the commodity sector as volatility in the commodity markets, agricultural in particular, are often related through natural causes. This would be an important area where such swaps would be useful for hedging risk. 

%\begin{acknowledgements}
%\end{acknowledgements}



\end{document}